\documentclass[a4paper,12pt]{article}
\usepackage{epsfig}
\usepackage{amssymb}
\usepackage{axodraw}

\newcommand{\nappend}{\setcounter{equation}{0}
\def\theequation{\rm{A}.\arabic{equation}}\section*}

\newcommand{\nnappend}{\setcounter{equation}{0}
\def\theequation{\rm{B}.\arabic{equation}}\section*}

\newcommand{\be}{\begin{equation}}
\newcommand{\ee}{\end{equation}}
\newcommand{\br}{\begin{eqnarray}}
\newcommand{\er}{\end{eqnarray}}
\newcommand{\ba}{\begin{array}}
\newcommand{\ea}{\end{array}}
\newcommand{\bi}{\begin{itemize}}
\newcommand{\ei}{\end{itemize}}
\newcommand{\bn}{\begin{enumerate}}
\newcommand{\en}{\end{enumerate}}
\newcommand{\bc}{\begin{center}}
\newcommand{\ec}{\end{center}}

\newcommand{\sw}{\sin\theta_W}
\newcommand{\sww}{\sin^2\theta_W}

\newcommand{\squaresm}{\lower0.085ex\hbox{$\square$}}

\newcommand{\gsim}{\lower.7ex\hbox{$\;\stackrel{\textstyle>}{\sim}\;$}}
\newcommand{\lsim}{\lower.7ex\hbox{$\;\stackrel{\textstyle<}{\sim}\;$}}

\newcommand{\plb}[3]{Phys.\ Lett.\ {\bf B#1} (19#2) #3}

\newcommand{\npb}[3]{Nucl.\ Phys.\ {\bf B#1} (19#2) #3}
\newcommand{\prd}[3]{Phys.\ Rev.\ {\bf D#1} (19#2) #3}

\newcommand{\prl}[3]{Phys.\ Rev.\ Lett.\ {\bf #1} (19#2) #3}

\def\sq{\tilde{q}}

\def\ar{\to}

\begin{document}
\tolerance=100000
\thispagestyle{empty}
\setcounter{page}{0}

\begin{flushright}
{\rm RAL-TR-1999-042}\\
{\rm TSL/ISV-99-0218}\\
{\rm September 1999} \\
\end{flushright}

\vspace*{\fill}

\begin{center}
{\Large \bf Effects of CP-violating phases}\\[0.5cm]
{\Large \bf on Higgs boson production at hadron colliders}\\[0.5cm]
{\Large \bf in the Minimal Supersymmetric Standard Model}\\[2.cm]

{{\Large A. Dedes}$^1$ {\large and} {\Large S. Moretti}$^{1,2}$} \\[5mm]
{\it $^1$Rutherford Appleton Laboratory,\\
Chilton, Didcot, Oxon OX11 0QX, UK} \\[5mm]
{\it $^2$Department of Radiation Sciences,
Uppsala~University, P.O.~Box~535,~75121~Uppsala,~Sweden} \\[5mm]
\end{center}

\vskip4.0cm

\begin{abstract}{\noindent
If the soft Supersymmetry (SUSY) breaking masses and couplings are complex,
then the associated CP-violating phases can in principle modify the known 
phenomenological pattern of the Minimal Supersymmetric Standard Model
(MSSM).   We investigate here their effects on Higgs boson production 
in the gluon-gluon fusion mode at the Tevatron and the Large Hadron Collider
(LHC),  by taking
into account all experimental bounds available at present.
The by far most stringent ones are those derived from the measurements
of the Electric Dipole Moments (EDMs) of fermions. However, it has
recently been suggested that, over a sizable  portion of the MSSM
parameter space, cancellations among the SUSY contributions to the
EDMs can take place, so that the CP-violating phases can evade those limits.  
We find a strong dependence of the production rates of any neutral Higgs
state upon the complex masses and couplings over such parts of the MSSM
parameter space. We show these effects 
relatively to the ordinary MSSM rates as well as illustrate them
at absolute cross section level at both colliders.
}\end{abstract}

\newpage

\section{Introduction and plan}
\label{sec:intro}

The soft SUSY breaking parameters of the MSSM can well be complex.
Even in the absence of  flavour non-conservation in the sfermion sector,
the Higgsino mass term, the gaugino masses, the trilinear couplings
and the  Higgs soft bilinear mass need not be real. 
Assuming universality of the soft gaugino masses at the Grand
Unification (GUT) (or Planck) scale,
the effects of complex soft masses and couplings in the
MSSM Lagrangian (see Appendix A) can be parametrised in terms
of only two independent phases \cite{hall}, $\phi_\mu$ and $\phi_A$,
associated to the (complex) Higgsino mass term, $\mu$, and to the
trilinear scalar
coupling $A$\footnote{For simplicity, hereafter (except in the Appendices), 
we assume $A\equiv A_u = A_d$ at the electroweak (EW)
scale, i.e., ${\cal O}(M_Z)$,
where $u$ and $d$ refer
 to all flavours of up- and down-type (s)quarks.}, respectively. 
In other terms,
\begin{equation}\label{phis}
e^{i\phi_\mu} \ =\ \frac{\mu}{|\mu|}\;, \;\;\;\;\;\;\;\;
e^{i\phi_A  } \ =\ \frac{A}{|A|} \;.
\end{equation}
Their presence is a potentially dangerous new source of violation of the 
CP-symmetry in the MSSM. But their size can in principle  strongly be
 constrained 
by the measurements of the fermionic EDMs (mainly, of electron and neutron)
and several analyses 
\cite{nir} have indicated that $\phi_\mu$ and $\phi_A$
 must be small in general. However, recent 
investigations  \cite{NATH,rosiek}  have shown that, in a restricted but still
sizable part of the parameter space of the MSSM, 
the bounds drawn from the EDM measurements 
 are rather weak, so that such phases 
can even be close to $\pi/2$. This is a consequence of cancellations
taking place among the SUSY loop contributions to the EDMs.
Although, in order to be effective, these require a certain amount of 
`fine-tuning' among the soft
masses and couplings \cite{rosiek}, it has recently 
been suggested that such cancellations occur naturally 
in the context of Superstring models \cite{brhlik1}.
If the SUSY loop contributions to the EDMs do vanish, then, 
as  emphasised by the authors of 
Ref.~\cite{Kane}, SUSY parameters with large imaginary parts
may have a non-negligible impact on the confrontation of the MSSM with  
experiments. In particular, many of the SUSY (s)particle production 
and decay processes 
develop a dependence on $\phi_\mu$ and $\phi_A$, so that,
in view of the importance of searches for New Physics 
at present and future accelerators,
their phenomenology needs a thorough re-investigation.

Various sparticle processes including the effect of such phases  
have recently been considered. For example, 
 neutralino  \cite{choi} and 
chargino \cite{herbi} production at LEP,
at the CERN LHC \cite{ma} as well as at future electron-positron
linear colliders (LCs) \cite{lai,barger}. Direct
CP-asymmetries in decays of heavy hadrons, 
such as $B\rightarrow X_s +\gamma$, $B\rightarrow X_d +\gamma$ and
 $B\rightarrow X_s l^+ l^-$, 
have been investigated in the context of the Supergravity inspired
MSSM (M-SUGRA): in Refs. \cite{aoki}, \cite{asatrian} and \cite{Huang},
respectively.

In this paper, we are concerned with the Higgs sector of the MSSM. 
Here, although the tree-level Higgs potential
is not affected by the CP-violating phases, since
$M_{H_1}^2$, $M_{H_2}^2$ and $\tan\beta$ 
(the mass parameters of the two Higgs fields
and the ratio of their vacuum expectation values (VEVs), respectively)
are real and $\mu$ enters only
through $|\mu|^2$, it should be noticed that
 this is no longer the case if one
includes radiative corrections. In their presence,
 one finds \cite{demir,pil}
that the three neutral Higgs bosons can mix and that their
effective couplings to fermions can be rather different at one loop.
However, for the MSSM parameter space that we 
will consider here, such corrections
turn out to be negligible, of the order of just a few percent,
as compared to those induced at the lowest order by the CP-violating 
phases in the squark sector. There are some reasons for this.
First of all, the induced radiative corrections to the Higgs-quark-quark
vertices
can be parametrised in terms of the mass of the charged Higgs boson, 
$M_{H^\pm}$ ($\approx M_{A^0}$). Then, one can verify that they essentially 
 depend  only upon the input values given to $|\mu|$, $|A|$
and $M_{\mathrm{SUSY}}$ (the typical mass scale of the SUSY partners of 
ordinary matter). Here, we will mainly be concerned with trilinear couplings
in the range  $|A| \lsim 700$ GeV, Higgsino masses $|\mu|$ of the order of
$600$ GeV or so, and $M_{\mathrm{SUSY}}\simeq 300$ GeV.
According to the analytic formulae of Ref.~\cite{pil}, in the above 
MSSM regime, one finds negligible corrections to the tree-level 
$h^0t\bar{t}$ and $h^0b\bar{b}$ couplings. (Similarly, for the 
case of the lightest Higgs boson mass.) In contrast, the strength of the  
$H^0t \bar{t}$ vertex can significantly be modified for not too heavy
masses of the charged Higgs boson (say, $M_{H^\pm}\approx M_{A^0} < 200$ GeV)  
and rather large values of $|A|$ and $|\mu|$ (typically, 
$|A|\simeq |\mu| \simeq 2$ TeV),
a region of parameter space that we will avoid, 
whereas that of 
the $H^0 b \bar b$ one is generally
small because we shall limit ourselves
to the interval $2 \lsim \tan\beta \lsim 10$. 
As for papers allowing  instead for the presence of non-zero values of 
$\phi_\mu$ and/or $\phi_A$ and studying the Higgs sector,
one can list Refs.
 \cite{choi2} and \cite{gunion}. The first publications deals with 
decay rates whereas the second one with Higgs production channels probable at a
future LC. In such papers though, no systematic treatment of the  
limits imposed by the EDM measurements was addressed. 
Such effects ought to be incorporated
in realistic analyses of the MSSM Higgs dynamics. 

Here, we have necessarily done so,  since
it is our purpose to study the effect of finite
values of $\phi_\mu$ and/or $\phi_A$ on
 Higgs production via the $gg\to \Phi^0$ channel \cite{xggh}, 
where $\Phi^0=H^0,h^0$ and $A^0$
represents any of the three neutral Higgs bosons of the MSSM. 
(A preliminary account in this respect was already given by the authors 
in Ref. \cite{prl}.) These processes  proceed
through quark (mainly top and bottom: i.e., $t$ and $b$, see 
Fig. 1) and squark 
(mainly stop and sbottom: i.e., ${\tilde t}_1,{\tilde t}_2$ and 
${\tilde b}_1,{\tilde b}_2$, see Fig. 2, 
each in increasing order of mass) loops, in which the (s)fermions
couple directly to $\Phi^0$.  Needless
to say, as in the MSSM the lightest of the Higgs particles
is bound to have a mass not much larger than that of the $Z$ boson, $M_Z$,
much of the experimental effort at both the Tevatron (Run 2) and the LHC
will be focused on finding this Higgs state, $h^0$.
In this respect, we remind the reader that direct Higgs production
via gluon-gluon fusion is the dominant mechanism over a large
portion of the MSSM parameter space at the LHC and a sizable
one at the Tevatron \cite{Higgsreview}. 

The plan of the paper is as follows. The next Section describes our theoretical
framework and discusses experimental limits on the parameters
of the model. The following one briefly sketches the way we have performed
the calculation. Sect.~\ref{sec:results} presents some numerical results, 
whereas Sect.~\ref{sec:conclusions} summarises our analysis and 
draws the conclusions. Finally, in
the two Appendices, we introduce our notation and explicitly derive 
the Feynman rules and cross section formulae needed for our numerical analysis.

\section{The theoretical model and its parameters}
\label{sec:parameters}

We work in the theoretical framework provided by  the MSSM, the
latter including explicitly
the  CP-violating phases and assuming universality of the soft gaugino
masses at the GUT scale and universality of the soft  trilinear
couplings at the EW scale. 
We define its parameters at the EW scale,
without making any assumptions about the structure of the
SUSY breaking dynamics
at the Planck scale, whether driven by Supergravity (SUGRA),  
gauge mediated (GMSB) or proceeding via other (yet unknown) mechanisms.
We treat the MSSM as a low-energy effective theory, and input all 
parameters needed for our analysis independently from each other. However,
we require these to be consistent with current experimental bounds. In fact,
given the dramatic impact that the latter can have on the viability at the 
Tevatron and/or the LHC
of the CP-violating effects in the processes we are dealing with,
we specifically devote the two following Subsections to discuss all available
experimental constraints. The first focuses on collider data, from 
LEP and Tevatron; the second on the measurements of the fermionic EDMs.
 (Some bounds can also be derived
from the requirement of positive definiteness of the  squark masses squared.)  
Following this discussion, we will
establish the currently allowed ranges for the Higgs and sparticle masses
and couplings.

Before proceeding in this respect though, we
declare the numerical values adopted for those MSSM parameters that 
are  in common with the Standard Model (SM).
For the top and bottom masses entering the SM-like fermionic
loops of our process, we have used 
$m_t=175$ GeV and $m_b=4.9$ GeV, 
respectively. As for the gauge couplings, 
the strong, electromagnetic (EM) coupling constants and the sine 
squared of the Weinberg angle, we have adopted
the following values:
$\alpha_s(M_Z)=0.119$,
$\alpha_{\mathrm{EM}}(M_Z)=1/127.9$
and $\sww(M_Z)=0.2315$, respectively.

\subsection{Limits from colliders}
\label{subsec:colliders}

The Higgs bosons and sparticles of the MSSM 
that enter the $gg\to \Phi^0$ production
processes can also be produced via other channels, both as real and
virtual objects. From their search at past and present colliders, 
several limits on their masses and couplings have been drawn. 
As for the neutral Higgs bosons of the MSSM,
the most stringent bounds come from LEP. For both $M_{h^0}$ and $M_{A^0}$
these are set -- after the 1998 LEP runs --  
at around 80 GeV by all Collaborations \cite{Kopp}, for
$\tan\beta>1$\footnote{See also 
\cite{LEP} for a more recent and somewhat higher limit
-- that we adopt here --
on $M_{h^0}$ from ALEPH, of about 85.2 GeV for 
$\tan\beta\ge1$ at 95\% confidence level (CL), using data collected at 
collider centre-of-mass (CM) energies in the range 
192 GeV $\lsim\sqrt s_{ee}\lsim$ 
 196 GeV and a total luminosity of about 100 pb$^{-1}$.}. The tightest
experimental limits on the squark masses come from direct searches at the
Tevatron. Concerning the ${{\tilde t}_1}$ mass,
for the upper value of
$\tan\beta$ that we will be using here, i.e., 10, the
limit on $m_{{\tilde t}_1}$ can safely be drawn at 120 GeV or so
\cite{Tevatron}, fairly independently of the SUSY model assumed. 
As for the lightest sbottom mass, $m_{{\tilde b}_1}$, this is excluded
for somewhat lower values, see Ref.~\cite{d0}.
Besides, D\O\ also contradicts
all models with $m_{\tilde{q}\ne\tilde{t}_1,\tilde{b}_1
}<$ 250 GeV for $\tan\beta\lsim2$, $A=0$ and
 $\mu< 0$  \cite{d01} (in scenarios with equal squark and gluino masses 
the limit goes up to $m_{\tilde{q}\ne\tilde{t}_1,\tilde{b}_1}<260$ GeV). 

\subsection{Limits from the EDMs}
\label{subsec:EDMs}

These are possibly the most stringent experimental constraints available 
at present on the size of the CP-violating phases. The 
name itself owns much to the consequences induced in the QED sector.
In fact, to introduce a complex part into the soft SUSY breaking
parameters of the MSSM 
corresponds to `explicitly' violating CP-invariance in the matrix elements
(MEs) involving the EM current, as the
phases lead to non-zero TP form factors, which in turn
contribute to the fermionic EDMs. In contrast, within the
SM, it is well known that contributions to the EDMs arise only 
from higher-order CP-violating effects in the quark sector,
and they are much smaller than the current experimental upper
bounds. At 90\% CL, those
 on the electron, $d_e$ \cite{de}, and
neutron, $d_n$ \cite{dn}, read as:
\begin{equation}\label{edmbounds}
|d_e|_{\rm{exp}} \le 4.3 \times 10^{-27}\; e \, {\rm cm} \;,
\qquad\qquad\qquad
|d_n|_{\rm{exp}} \le 6.3 \times 10^{-26} \; e \, {\rm cm} \;. 
\end{equation}
As mentioned in the Introduction, if cancellations take place 
among the SUSY contributions to the electron and neutron EDMs, so that 
their value in the MSSM is well below the above limits, i.e.,
$|d_e|_{\rm{MSSM}}\ll|d_e|_{\rm{exp}}$
and
$|d_n|_{\rm{MSSM}}\ll|d_n|_{\rm{exp}}$, then $\phi_\mu$ and
$\phi_A$  can be large.
To search for
those combinations of soft sparticle masses and couplings that guarantee 
vanishing SUSY contributions to the EDMs for each possible choice of the
CP-violating phases, we have scanned over the  
($\phi_\mu, \phi_A$) plane and made use of the program
of Ref.~\cite{rosiek}. This returns those minimum values of
the modulus of the common trilinear coupling, $|A|$, 
above which the cancellations work.
For instance, in the case of the neutron EDMs, 
the dominant  chargino and gluino contributions appear 
with opposite sign over a large portion of the MSSM parameter space.
Thus, for a given $|\mu|$, a chargino diagram  can cancel a
gluino one and this occurs for certain values of the
gaugino/squark masses and a specific choice of $|A|$ \cite{NATH,rosiek}.
In general, internal  cancellations are more likely among the SUSY
 contributions to the neutron EDMs, than they are in the case of the electron.
So much so that, in the former case,
it is even possible to remain consistently above the experimental limits
(\ref{edmbounds})
if one only assumes the phase of $|\mu|$ to be non-zero~\cite{rosiek}.

However, not all the surviving combinations of $\phi_\mu$, $\phi_A$ and
$|A|$ are necessarily allowed. In fact,
one should recall that physical parameters of the MSSM
depend upon these three inputs. In particular,
the squark masses (entering the triangle loops of the production
processes considered here) are strongly related to 
$\phi_\mu$, $\phi_A$ and
$|A|$. Given the assumptions 
already made on the trilinear couplings (i.e., their universality), 
and further setting (see Appendix A for the notation)  
\br
M_{\tilde{q}_3}\equiv M_{\tilde{Q}_3}=M_{\tilde{U}_3}=M_{\tilde{D}_3}\;, 
\\[2mm]
M_{\tilde{q}_{1,2}} \equiv M_{\tilde{Q}_{1,2}}=M_{\tilde{U}_{1,2}}=
M_{\tilde{D}_{1,2}} \;,
\er
with $M_{\tilde{q}_{1,2}} \gsim M_{\tilde{q}_3}$,
where $M_{\tilde{q}_{1,2,3}}$ are the soft squark masses of the 
three generations, one gets for the lightest stop and sbottom masses
the following relations:
\br 
m^2_{\tilde{t}_1} \ &=& \ M_{\tilde{q}_3}^2 + m_t^2 +
\frac{1}{4}M_Z^2\cos2\beta - \nonumber \\ & & \sqrt {(\frac{5}{6}M_Z^2-
\frac{4}{3}M_W^2)^2\cos^2 2\beta + 4 m_t^2\biggl [|A|^2 +
|\mu|^2 \cot^2\beta
 + 2 |A| |\mu| \cos(\phi_\mu - \phi_A) \cot\beta \biggr] } \;, \nonumber \\ 
\label{lst} \\
m^2_{\tilde{b}_1} \ &=& \ M_{\tilde{q}_3}^2 +m_b^2 - 
\frac{1}{4}M_Z^2\cos2\beta - \nonumber \\ & & \sqrt{(\frac{1}{6}M_Z^2-
\frac{2}{3}M_W^2)^2\cos^2 2\beta + 4  m_b^2 \biggl [ |A|^2 +
|\mu|^2 \tan^2\beta 
+ 2 |A| |\mu| \cos(\phi_\mu - \phi_A) \tan\beta \biggr ]} \;.  \nonumber \\
\label{lsb}
\er
For some choices of $\phi_\mu$, $\phi_A$ and $|A|$ and a 
given value of $M_{\tilde{q}_3}$, $|\mu|$ and $\tan\beta$, 
the two above masses (squared) can become
 negative. This leads
to a breaking of the SU(3) symmetry, that is, to the appearance of colour and 
charge breaking minima. In order to avoid this, some points on the plane
($\phi_\mu,\phi_A$) will further be excluded in our study.

\section{Numerical calculation}
\label{sec:calc}

We have calculated the Higgs production rates in presence of the CP-violating
phases exactly at the leading order (LO) and compared them to the
yield of the ordinary MSSM (that is, `phaseless')   
at the same accuracy. In our simulations, we have included only the 
$t$-, $b$-,  ${\tilde t}_1$-, ${\tilde t}_2$-, 
${\tilde b}_1$- and ${\tilde b}_2$-loops, 
indeed the dominant terms, because of the Yukawa type couplings involved. 
In order to do so, we had to compute from scratch all the
relevant analytical formulae for $A^0$ production. In fact, 
one should notice that for such a Higgs state there exist no tree-level 
couplings with identical squarks if $\phi_\mu=\phi_A=0$, whereas they appear
at lowest order whenever one of these two parameters is non-zero. Besides,
being the quark loop contributions antisymmetric (recall that 
$A^0$ is a pseudoscalar
state) and the squark loop ones symmetric, no interference effects can take
place between the SM- and the SUSY-like terms in the ME
for $gg\to A^0$. (That is to say that $\phi_\mu$- and $\phi_A$-induced
corrections are always positive if $\Phi^0=A^0$.)
The full amplitude is given explicitly in Appendix B.
For completeness, we have also recomputed the well known expressions
for scalar Higgs production, $\Phi^0=h^0,H^0$, finding perfect agreement
with those already given in literature (again, see Appendix B).
Here, CP-violating effects can produce corrections of both signs.

It is well known that next-to-leading order (NLO) corrections to
$gg\to\Phi^0$ processes 
from ordinary QCD are very large \cite{corr,DDS}. However,
it has been shown that they affect the quark and squark contributions
very similarly \cite{DDS}. Thus, as a preliminary exercise, one can look 
at the LO rates only in order to estimate the effects induced
by the CP-violating phases. In contrast, for more phenomenological analyses, 
one ought to incorporate these QCD effects. We have eventually
done so by resorting to the analytical expressions for the heavy 
(s)quark limit given in Ref.~\cite{DDS}. These are expected to be a very good
approximation for Higgs masses below the quark-quark and squark-squark 
thresholds. Therefore, we will confine ourselves to 
combinations of masses which respect such kinematic condition. 

As Par\-ton Dis\-tri\-bu\-tion Func\-tions (PDFs), 
we have used the fits MRS98-LO(05A) \cite{twentythree} and
MRS98-NLO(ET08) \cite{MRS98}, in correspondence of our one- and two-loop
simulations, respectively. Consistently, we have adopted the one- and two-loop 
expansion for the strong coupling constant $\alpha_s(Q)$, with all relevant 
(s)particle thresholds onset within the MSSM as described in \cite{Dedes}.
The running of the latter, as well as the evolution of the PDFs, was always
described in terms of the factorisation scale $Q\equiv Q_F$, which was
set equal to the produced Higgs mass,
$M_{\Phi^0}$. In fact, 
the same value was adopted for the renormalisation scale 
$Q\equiv Q_R$ entering the Higgs production processes, see eqs.~(\ref{me})
and (\ref{oddme})--(\ref{evenme}).

Finally, for CM energy of the LHC, we have assumed
$\sqrt s_{pp     }=14$ TeV; whereas for the Tevatron, we have taken 
$\sqrt s_{p\bar p}=2$ TeV.

\section{Results}
\label{sec:results}

A sample  of $\phi_\mu$, $\phi_A$ and $|A|$ values that guarantee
the mentioned cancellations can be found in 
Fig.~\ref{fig:A}, for the two representative choices of
$M_{{\tilde q}_{1,2}}$ given in 
Tab. I. Here, both $|\mu|$ and $M_{{\tilde q}_{3}}$ are held constant at,
e.g., 600 and 300 GeV, respectively. 
The soft gluino mass is given too, in Tab.~I, as it enters our analysis
indirectly, through the EDM constraints (recall the discussion in 
Subsect.~\ref{subsec:EDMs}).
The allowed values for the modulus of the (common) trilinear couplings $|A|$
are displayed in the form of a contour plot over the ($\phi_\mu,\phi_A$)
plane, where both
 phases  are varied from $0$ to $\pi$. (Same results are obtained in the
interval $(\pi, 2\pi)$, because of the periodic form of the SUSY couplings
and mixing angles: see Appendix A.)
In the same plots, 
we have superimposed those regions 
(to be excluded from further consideration)
over which the observable MSSM parameters assume values that are either
forbidden by collider limits (dots for
the lightest stop mass and squares for the lightest Higgs mass:
 see Figs.~\ref{fig:stop1}--\ref{fig:higgs} below)
or for which the squark masses  squared
become negative (crosses), for a given 
combination of the other soft SUSY breaking parameters.
Typically, we obtain that for small $\mu$ phases, i.e., 
$\phi_\mu\lsim \pi/300\approx 0.01$, 
 the value of  $|A|$ tends to be zero for almost 
all values of $\phi_A$.  In the region where both phases are quite large
($\phi_\mu,\phi_A\approx\pi/2$), the modulus of
the trilinear coupling must be around  700 GeV, for
the EDM constraints to be satisfied.

In Tab.~I, in order to completely define our model 
for the calculation of 
the $gg\to \Phi^0$ processes, we also have introduced 
the Higgs sector parameters:  the  mass of one  
physical states, e.g., $M_{A^0}$, and
the ratio of the VEVs of the two doublet fields, i.e., 
$\tan\beta$. We have fixed the former to be 200 GeV, whereas two
 possible choices of the latter have been adopted, $2.7$ and
$10$ (corresponding to $M_{\tilde{q}_{1,2}}=1000$ and 300 GeV, 
respectively). With regard to this last parameter, 
$\tan\beta$, a few
considerations have to be made at this point: namely, concerning
 the so-called `Barr-Zee type graphs' \cite{zee}.
Very recently, the corresponding contributions to the
electron and neutron EDMs  have been calculated at two loops, in Ref.
\cite{chang}. These terms put bounds (diamond symbols) directly on the
squark masses and soft trilinear couplings of the third generation
and thus are crucial for our  analysis. However, they are quantitatively
 significant only at large values of $\tan\beta$, as can be
appreciated in the right-hand plot of Fig.~\ref{fig:A}. The smaller
$\tan\beta$ the less relevant they are: see the left-hand plot of 
the same figure. Besides, these two-loop terms entering 
the EDMs also depend on $|\mu|$: small values of the latter induce 
negligible contributions to both the electron and neutron EDMs.
(For example, for the choice $|\mu|=500$ GeV and $\tan\beta=3$
made in Ref. \cite{prl}, one obtains
no bounds from the EDMs through the Barr-Zee type graphs.)  Moreover, 
for $\tan\beta\gsim10$, in order to be consistent 
with the EDM constraints in (\ref{edmbounds}), one would
need the modulus of the soft trilinear coupling to be unnaturally large, even
greater than $4$ TeV  (for $\phi_A\approx\phi_\mu\approx\pi/2)$.
This would drive the squark masses in eqs.~(\ref{lst})--(\ref{lsb}) 
to become negative over most of the ($\phi_\mu$,$\phi_{A}$) plane. Thus, 
in order to avoid all such effects, we limit 
our analysis to the interval $2\lsim\tan\beta\lsim10$ 
(recall the lower experimental limit on such a parameter: see
Subsect.~\ref{subsec:colliders}). 
Corresponding values for the modulus of the (common) trilinear couplings are
in the range $|A|\lsim 1$ TeV.
In this regime, 
the parameter combinations given
in Tab.~I should serve the sole purpose 
of being  examples of the rich phenomenology that can be
induced by the CP-violating phases in the MSSM, rather than
benchmark cases. 

\begin{center}
\begin{tabular}{|c|c|c|c|c|c|}
\hline
$\tan\beta$ & $|\mu|$ & $M_{\tilde{q}_{1,2}}$ & $M_{\tilde{q}_3}$  
& $M_{\tilde{g}}$ & $M_{A^0}$  \\[2mm] \hline 
2.7  & 600 & 1000 & 300 & 300 & 200  \\ \hline 
10 & 600 & 300 & 300 & 300 & 200  \\ \hline 
\end{tabular}

\vspace*{2mm}

{\small Table I:  Two
possible parameters setups of our model.
(Apart from the dimensionless $\tan\beta$, all other
quantities are given in GeV.)}
\end{center}

Furthermore, notice that, starting from the numbers in Tab. I, 
one can verify that the heaviest squark masses, 
$m_{\tilde{t}_2}$ and $m_{\tilde{b}_2}$,
are both 
consistent with current experimental bounds. As for the lightest stop, 
we display in Fig.~\ref{fig:stop1} the values assumed 
by $m_{\tilde{t}_1}$
over the usual ($\phi_\mu,\phi_A$) plane, for the two choices
of $\tan\beta$ in Tab.~I and in correspondence
of the $|A|$ values of the previous  figure. As a matter of fact, 
the effect of the phases is quite 
significant on the actual lightest stop mass, {see eq.~(\ref{lst})}.
We observe that, for $\phi_\mu\approx\frac{\pi}{2}$, one 
can get small values for $m_{{\tilde t}_1}$, if not negative. 
This is due to the fact that, in such a 
region, $|A|$ can still get large enough (despite of a
low $\tan\beta$), so that the last term on the
right-hand side of  eq.~(\ref{lst})
becomes comparable to the first two terms.
Anyhow, over most of the ($\phi_\mu,\phi_A$)
plane, $m_{\tilde{t}_1}$ is well above
the current experimental reach of 120 GeV.
The lightest sbottom mass, $m_{{\tilde b}_1}$, see
eq.~(\ref{lsb}), is always around
  $290$ GeV, so that we have avoided to reproduce here the corresponding
plots.  
Also, for the above choices of $\tan\beta$, one gets
constraints on the mass of the lightest Higgs boson. 
In order to derive these, we
make use of the two-loop analytic formula for $M_{h^0}$  \cite{Hollik}.
We display the corresponding values
over the   ($\phi_\mu,\phi_A)$ plane in Fig.~\ref{fig:higgs}. The
 mass regions excluded 
by LEP (see Subsect.~\ref{subsec:colliders})
amount to a restricted part of the ($\phi_\mu,\phi_A$) plane, both at small
and, particularly, large $\tan\beta$. As for the heaviest
Higgs boson masses, one has $M_{H^0}=212 (201)$ GeV for
$\tan\beta=2.7 (10)$, thus almost degenerate with $M_{A^0}$.

Having defined the allowed spectrum for the (s)particles masses
and couplings entering the Higgs production modes that we are considering,
we  are  now ready to quantify the effects 
of the phases on the actual cross sections.
A convenient way of doing so is to simply look at the ratio
between the MEs computed with and without
phases. In fact, at LO accuracy and further assuming 
that the relevant hard scale is the same in both cases (e.g., 
$Q\equiv M_{\Phi^0}$), such a ratio coincides with that obtained
at cross section level, independently of the choice of the PDFs and of
$\sqrt s$.
Thus, we define 
\br
R(g g \ar \Phi^0) \ =\  
\frac{\sigma^{{\mathrm{MSSM}}^*}_{\mathrm{LO}}(g g \ar \Phi^0)}
     {\sigma^{{\mathrm{MSSM}}~ }_{\mathrm{LO}}(g g \ar \Phi^0)}\;.
\label{rat}
\er
By means of the notation 
MSSM$^*$, we refer here and in the
following to the case of the MSSM with same $|A|$ but 
finite values of either  $\phi_\mu$ or $\phi_A$. Thus, e.g., when 
$\phi_\mu=\phi_A=0$ (and $|A|=0$ too, see Fig.~\ref{fig:A}),
 the expression in (\ref{rat})
is of course equal to $1$. 
(Several results for this ratio have already been 
presented in Ref.~\cite{prl}, though
for a choice of the MSSM parameters different
from those considered here.)

In Fig.~\ref{fig:h0} we present
 $R(g g \ar h^0)$ as a contour plot over the  
($\phi_\mu,\phi_A$) plane for the two choices of MSSM parameters of Tab.~I.
As a consequence of the expression for the $gg\to h^0$ amplitude, see
eq.~(\ref{me}), the corrections induced by the
presence of finite values of $\phi_\mu$ and/or $\phi_A$ in the
squark loops can be either positive or negative. Interestingly enough,
destructive interferences take place over regions already 
 excluded by direct Higgs searches. Only constructive interferences
would then be observable. These can considerably 
enhance the value of the cross sections obtained in the 
ordinary MSSM, particularly 
at large $\phi_A$ and intermediate  $\phi_\mu$ values, if $\tan\beta$
is small (by up to a factor of 7). This
result is a consequence of two related aspects. Firstly, 
in those ($\phi_\mu,\phi_A$) regions, $|A|$ achieves its maximum,
so that the corresponding Higgs-squark-squark couplings 
are enhanced significantly, with respect to the strength of the
Higgs-quark-quark vertices (see Appendix A). 
Secondly, large values of $|A|$ correspond to small values of
$m_{{\tilde t}_1}$ (compare  Fig.~\ref{fig:A} to  Fig.~\ref{fig:stop1}),
this yielding a further kinematic enhancement in the squark loops.
For large $\tan\beta$, the portion of the ($\phi_\mu,\phi_A$)
plane surviving the EDM constraints is much smaller, but the 
CP-violating effects are
still large. For example, in the region 
$3\pi/5 \lsim \phi_\mu \lsim 3\pi/4$ 
and $\pi/3 \lsim \phi_A \lsim 3\pi/4$,
the MSSM$^*$ cross sections can be larger by about 
a factor of 5 with respect to the MSSM ones.

In Fig.~\ref{fig:H0} we present similar rates for the heaviest scalar
Higgs boson, i.e., $\Phi^0=H^0$. Contrary to the previous case,
here, one typically obtains a suppression
of the MSSM$^*$ cross sections relatively to the MSSM ones, over the allowed
regions of the $(\phi_\mu,\phi_A)$ plane, both at small and large
$\tan\beta$. The 
destructive effect is the consequence  of the interplay between the Higgs
mixing angle $\alpha$ and the minus sign  
in the third term of the relevant 
$\lambda_{H^0{{\tilde t}_1}{{\tilde t}_1^*}}$ coupling,
see eq.~(\ref{Ht1t1}), as opposed to a plus sign in 
 $\lambda_{h^0{{\tilde t}_1}{{\tilde t}_1^*}}$ of eq.~(\ref{ht1t1}).
Quantitatively, the effects of the phases are most conspicuous at
small $\tan\beta$, 
when $\phi_\mu \simeq 3\pi/5$  for small $\phi_A$ values (about one
order of magnitude difference between the MSSM$^*$ and the MSSM, at the most).
At large $\tan\beta$, over the much smaller ($\phi_\mu,\phi_A$) plane
surviving the experimental constraints, the suppression is at most
a factor of 2, when  $\phi_\mu \simeq 4\pi/5$.

The ratio $R(g g \ar A^0)$ is plotted in Fig. \ref{fig:A0}.
As already discussed in Sect.~\ref{sec:calc}, here the
corrections induced by finite values of $\phi_\mu$ and/or $\phi_A$ are
always positive: see  eq.~(\ref{me}).
In fact, the ratio    $R(g g \ar A^0)$ can become as large as 2
at small $\tan\beta$. In this case, 
the maximum is obtained very close to the excluded regions: i.e.,
$\phi_\mu \simeq \pi/2$ and $\phi_A \simeq 0,\pi$. This
pattern can easily be understood by looking at the  
$\lambda_{A^0\tilde{t}_1 \tilde{t}_1^*}$ coupling of 
eq.~(\ref{At1t1}) and further recalling eq.~(\ref{ang}). 
(Also notice that, for 
$\phi_A=\phi_\mu \simeq \frac{\pi}{2}$ and $\tan\beta =2.7$,
the coupling $\lambda_{A^0\tilde{t}_1
\tilde{t}_1^*}$ becomes zero, hence $R(gg\to A^0)=1$.)
For larger values of $\tan\beta$, the effects
of the phases on pseudoscalar Higgs boson production
 can even be larger, as they are induced through
the  $ A^0\tilde{b}_1 \tilde{b}_1^*$ vertex
of eq.~(\ref{At1t1}), which benefits from the  $\tan\beta$ enhancement.  
For example, when $\phi_\mu\simeq 3\pi/4$ and for $\phi_A$ slightly
larger than $\pi/2$,  the increase can amount to a factor of 3.

As general remark on the behaviour of
 the three ratios, one should notice that they are
close to unity (i.e., no effects from the CP-violating phases) when 
$\phi_\mu$ is small for every value of $\phi_A$. This can 
easily be interpreted by looking at Fig. \ref{fig:A}, 
since, when $\phi_\mu \ar 0$,
one has that $|A| \ar 0$ and $\phi_{\tilde{t},\tilde{b}} \ar 0$ too (these 
are the mixing
angles of the third generation, see Appendix A).
Thus, the strength of all Higgs-squark-squark vertices becomes
 very small compared to that of the Higgs-quark-quark ones. The opposite 
($|A| \ar 0$  when $\phi_A\to0$ for any $\phi_\mu$) is not true, since 
here the 
$|\mu|$ value is fixed and thus $\phi_{\tilde{t},\tilde{b}}$ 
are always non-zero.

As already intimated in Sect.~\ref{sec:calc}, in order to give realistic
predictions for CP-violating effects in $gg\to\Phi^0$ processes, 
one ought to include
two-loop QCD effects. We do so in the reminder of this Section, by
considering the NLO production rates of all Higgs states at
the mentioned CERN and FNAL colliders,  
$\sigma_{\mathrm{NLO}}^{\mathrm{MSSM}^*}(gg \ar \Phi^0)$. 
 We convert the total production
cross sections to picobarns and again adopt the input parameters of  Tab.~I. 
 (Incidentally,
notice that, being $m_{{\tilde{t}}_1}>120$ GeV, the relation 
 $M_{\Phi^0} < 2 m_{\tilde{q}}$ is always satisfied, for any
$\tilde q=\tilde{t}_1,\tilde{t}_2,\tilde{b}_1,\tilde{b}_2$: recall the
discussion in
Sect.~\ref{sec:calc}.) The LHC rates are displayed through 
Figs.~\ref{fig:sigmah0LHC}--\ref{fig:sigmaA0LHC}
whereas the Tevatron ones appear in 
Figs.~\ref{fig:sigmah0TEV}--\ref{fig:sigmaA0TEV}.
We present these figures mainly as a reference for experimental analyses.
In fact, as far as CP-violating effects are concerned,
the two-loop QCD dynamics is very similar to the one-loop one
already discussed. In particular, we have verified that the QCD
$K$-factors
for the quark and squark loops are very similar \cite{DDS} over the portions
of the MSSM parameter space considered here. Thus, to obtain the effects
of $\phi_\mu$ and $\phi_A$ on Higgs production via gluon-gluon fusion at
NLO, it suffices to refer to Figs.~\ref{fig:h0}--\ref{fig:A0}
with the normalisation of Figs.~\ref{fig:sigmah0LHC}--\ref{fig:sigmaA0LHC}
and Figs.~\ref{fig:sigmah0TEV}--\ref{fig:sigmaA0TEV}, for the LHC and 
the Tevatron, respectively.
Concerning the possibility of actually observing the CP-violating effects
at either collider, this is very much dependent upon their luminosities. 

At the LHC, with an annual value between
10 and 100 fb$^{-1}$, all the available ($\phi_\mu,\phi_A$) 
areas  can in principle 
be covered, at both large and small $\tan\beta$ and for all Higgs
states, as the production cross sections are never smaller than
5 pb or so, see Figs.~\ref{fig:sigmah0LHC}--\ref{fig:sigmaA0LHC}. 
In particular, this is true where the 
effects of $\phi_\mu$ and $\phi_A$ are larger: compare the areas with
high $R(gg\to\Phi^0)$ values in Figs.~\ref{fig:h0}--\ref{fig:A0} to
the corresponding rates in Figs.~\ref{fig:sigmah0LHC}--\ref{fig:sigmaA0LHC}.
For example, for $\Phi^0=h^0$, at large $\phi_A$, intermediate
$\phi_\mu$ and for $\tan\beta=2.7$,
$\sigma_{\mathrm{NLO}}^{\mathrm{MSSM}^*}(gg \ar h^0)$
is around 200 pb.  At large $\tan\beta$, when
$3\pi/5 \lsim \phi_\mu \lsim 3\pi/4$  
and $\pi/3 \lsim \phi_A \lsim 3\pi/4$,
the MSSM$^*$ cross sections are around 100 pb or more.
Similarly, for $\Phi^0=H^0$, if $\tan\beta=2.7$, 
when $\phi_\mu \simeq 3\pi/5$  and for small $\phi_A$ values,
$\sigma_{\mathrm{NLO}}^{\mathrm{MSSM}^*}(gg \ar H^0)$
is of the order of 6 pb.
At $\tan\beta=10$,  when  $\phi_\mu \simeq 4\pi/5$ (for any $\phi_A$),
one gets cross section rates around 10 pb. Finally, if $\Phi^0=A^0$,
at small $\tan\beta$ and when
$\phi_\mu \simeq \pi/2$ with $\phi_A \simeq 0,\pi$, one
has again MSSM$^*$ cross sections of the order of 10 pb.
For large $\tan\beta$, $\phi_\mu\simeq 3\pi/4$ and $\phi_A\gsim
\pi/2$, $\sigma_{\mathrm{NLO}}^{\mathrm{MSSM}^*}(gg \ar A^0)$
is about 15 pb.

Concerning the Tevatron, given the value of integrated luminosity expected
at Run 2, of the order of 10 fb$^{-1}$,
prospects of detecting CP-violating effects in $gg\to \Phi^0$
processes are very slim.
The various $\sigma_{\mathrm{NLO}}^{\mathrm{MSSM}^*}(gg \ar \Phi^0)$'s
are notably smaller here, because of the reduced gluon content
inside the (anti)proton at lower $\sqrt s$, for a given $M_{\Phi^0}$
value. In fact, the production rates over not yet excluded ($\phi_\mu,\phi_A$)
regions are never larger than a handful of picobarns. One can possibly
aim at disentangling  CP-violating effects in the case of the lightest
Higgs boson, at both $\tan\beta$'s,  
but only in the usual corners where
both the corrections and the absolute production rates are largest:
see Fig.~\ref{fig:sigmah0TEV}. As for the
other two Higgs states, the chances are extremely poor, as 
$\sigma_{\mathrm{NLO}}^{\mathrm{MSSM}^*}(gg \ar \Phi^0)$ is always 
well below the picobarn level, see 
Figs.~\ref{fig:sigmaH0TEV}--\ref{fig:sigmaA0TEV}.  

As a final remark of this Section, 
we would like to mention the following.
A peculiar feature concerning Figs.~\ref{fig:sigmah0LHC}--\ref{fig:sigmaH0LHC},
as compared to Figs.~\ref{fig:h0}--\ref{fig:H0}, respectively, is the
different pattern of the level curves. This should not be surprising though,
as, for $\Phi^0=h^0,H^0$, the squark loop contributions in the ordinary
MSSM are non-zero to start with. In contrast, one can appreciate the
strong correlations between the level curves in
Fig.~\ref{fig:sigmaA0LHC} and Fig.~\ref{fig:h0}, a consequence 
of the absence of scalar loops in $gg\to A^0$ if $\phi_\mu=\phi_A=0$.
Similarly, for the case of Figs.~\ref{fig:sigmah0TEV}--\ref{fig:sigmaA0TEV}.

\section{Summary and conclusions}
\label{sec:conclusions}

It is well known that finite values of the mixing
angles $\theta_{\tilde t,\tilde b}$,
converting the weak into the mass basis of the third generation
of squarks, see eqs.~(\ref{unitary}) and
(\ref{convert}), imply that left-right chiral currents,
eqs.~(\ref{LR})--(\ref{RL}), can enter the 
Higgs-squark-squark couplings appearing in the scalar
loops contributing to
 $gg\to \Phi^0$ processes, if $\Phi^0=h^0,H^0$, because of the
structure of the mixing equations (\ref{mixing}).
As a consequence, the corresponding $\Phi^0$ production
cross sections develop a dependence
on $\mu$ and $A$, the Higgsino mass term and the trilinear scalar
coupling (the latter assumed here to be universal to all (s)quark flavours)
entering the soft SUSY breaking sector. The strength of their contribution
 is however modified if
CP-violating effects are manifestly inserted into the MSSM Lagrangian, by 
allowing these two parameters to be 
complex, see eqs.~(\ref{At1t1})--(\ref{hb2b2}).
In such a case, in particular, also the
$gg\to A^0$ cross section receives scalar loop contributions, thus acquiring 
a dependence upon $\mu$ and $A$, much on the same footing as when
$\Phi^0=h^0,H^0$: see eq.~(\ref{me}).

Clearly, it is the actual size of the independent phases
associated to the above two parameters, $\phi_\mu$ and $\phi_A$, that
regulates the phenomenological impact of complex values of $\mu$ and/or $A$,
not least, because they also affect the two mixing angles: see 
eqs.~(\ref{tmix}) and (\ref{bmix}). 
Given the importance of Higgs production at future hadron-hadron colliders,
such as the Tevatron (Run 2) and the LHC, and the fact that gluon-gluon
fusion is a sizable production mode at the former and indeed the dominant
one at the latter (over most of the MSSM parameter space), we have 
made the investigation of the effects induced by finite CP-violating phases
in $gg\to\Phi^0$ processes the concern of this paper. In order to address
the problem quantitatively, we first had to derive the relevant Feynman rules
of the MSSM in presence of $\phi_\mu$ and $\phi_A$
(Appendix A) and
eventually calculate the associated cross sections, for any Higgs state
(Appendix B).

Before proceeding to the numerical analysis though, we had
to introduce a parametrisation of our theoretical model and
incorporate the latest experimental constraints on its parameters. 
These can be subdivided into 
two categories, those arising from analyses performed with collider data 
and those deduced from the measurements
of the electron and neutron EDMs. The former mainly limit the
value of the squark masses and couplings, thus only indirectly affecting
$\phi_\mu$ and $\phi_A$, see eqs.~(\ref{lst})--(\ref{lsb}). In contrast,
the latter can be very stringent in this respect, unless
 cancellations
take place among the SUSY contributions to the fermionic EDMs, so that
the bounds derived this way on $\phi_\mu$ and $\phi_A$ 
become much less potent.
In practise, the CP-violating phases can attain any value between $0$ and
$\pi$, provided $|\mu|$ and $|A|$ are in appropriate relations, 
which are in fact satisfied
 over large portions of the MSSM parameter space. Under these 
circumstances then, the CP-violating phases can  affect the 
interplay between the quark and squark loops in $gg\to\Phi^0$ processes
considerably (Section 2).

In the end, we have verified, both at LO and NLO 
accuracy (Section 3), that this is true,
over those parts of MSSM parameter space where these cancellations
are more effective. As a matter of fact, effects due to finite values
of $\phi_\mu$ and/or $\phi_A$  can be extremely 
large, inducing variations on the Higgs cross sections of the ordinary
MSSM (i.e., those obtained 
for $\phi_\mu=\phi_A=0$
at the same $|A|$) of several hundred percent,
at least for values of $\tan\beta$ in the range between 2 and 10
and soft masses and couplings below the TeV region
 (Section 4). For these combinations
of parameters, even the bounds induced by the contributions
of the Barr-Zee type diagrams to the EDMs can easily be evaded.
Other than studying relative effects of the phases, with respect to the
yield of the ordinary MSSM, 
we also have presented absolute rates for the $gg\to \Phi^0$
 cross sections at NLO,
at both the LHC and the Tevatron, for the MSSM including $\phi_\mu$ and
$\phi_A$, thus showing CP-violating
 effects explicitly in observable quantities.
Given the higher luminosity and production rates expected at the CERN collider,
as compared to the values at the FNAL one, real prospects of sizing 
these effects in Higgs boson production will most likely have
to wait for a few more years. By then, the available
portions of the ($\phi_\mu,\phi_A$) plane should also be expected to
be better defined than at present, 
given the improvement foreseen in the near future in the precision
of the EDM measurements \cite{EDM}.

Anyhow, as
we have tried to motivate in the Introduction, and following our results,
we believe that further investigation is needed of the consequences of
explicit CP-violation being present
in the soft SUSY Lagrangian. For example, to stay
with the Higgs sector, one should establish 
the effects of $\phi_\mu$ and $\phi_A$  in the decay process 
$h^0\to\gamma\gamma$ \cite{decay}, as this represents the 
most promising discovery channel of the lightest Higgs boson 
of the MSSM at hadron-hadron machines.
In this case, the proliferation of SUSY induced contributions
(also due to charged Higgs bosons, sleptons and gauginos)
could well be responsible of CP-violating effects comparable to those
seen in the Higgs production processes via gluon-gluon fusion, as the latter
are solely due to squarks contributions.

\section*{Acknowledgements}

S.M. acknowledges  financial support from the UK PPARC and
A.D. that  from the Marie Curie Research Training Grant            
ERB-FMBI-CT98-3438. We thank J. Rosiek for providing us with 
his computer program and for his assistance in using it.
A.D. would also like to thank Dick Roberts, Mike Seymour and Apostolos 
Pilaftsis for helpful discussions. We both thank C. Schappacher
for spotting a typo in the first version of this manuscript.


\newpage
\nappend{Appendix A: the CP-violating phases in the MSSM}

In this Section, we follow the notation of Ref.~\cite{HHG}. We start from
the Superpotential, which has the form
\br
{\cal W} \ =\ \epsilon_{ij} \left ( Y_e H_1^i L^j \bar{E} +
Y_d H_1^i Q^j \bar{D} + Y_u H_2^j Q^i \bar{U} + \mu H_1^i H_2^j \right )
\er
and where all fields appearing are actually 
Superfields, with $\epsilon_{12}=1$. 
In terms of component fields, the Lagrangian of
the soft breaking terms reads as 
\br
{\cal L}_{soft} \ &=& \ -M_{H_1}^2 |H_1|^2 - M_{H_2}^2 |H_2|^2
-\mu B \epsilon_{ij} \left (H_1^iH_2^j + {\mathrm{h.c.}} \right ) \nonumber \\
&-&\frac{1}{2}M_1\bar{\tilde{B}}\tilde{B}-\frac{1}{2}M_2\bar{\tilde{W^a}} 
\tilde{W^a}-\frac{1}{2}M_3\bar{\tilde{g^\alpha}}\tilde{g^\alpha} 
\nonumber \\
&-&M_{\tilde{Q}}^2 \left (\tilde{u}_L^*\tilde{u}_L + \tilde{d}_L^*\tilde{d}_L
\right ) -M_{\tilde{U}}^2\tilde{u}_R^*\tilde{u}_R-
M_{\tilde{D}}^2\tilde{d}_R^*\tilde{D}_R \nonumber \\
&-&M_{\tilde{L}}^2 \left (\tilde{e}_L^*\tilde{e}_L + 
\tilde{\nu}_L^*\tilde{\nu}_L \right )
-M_{\tilde{E}}^2\tilde{e}_R^*\tilde{e}_R  \nonumber \\
&-& \epsilon_{ij} \left (-Y_u A_u H_2^i \tilde{Q}^j\tilde{u}_R^* +
Y_d A_d H_1^i\tilde{Q}^j \tilde{d}_R^* + Y_e A_e H_1^i \tilde{L}^j 
\tilde{e}_R^*  +{\mathrm{h.c.}} \right ) \;.
\er
The squark mass squared matrix (here and in the following, $q^{(')}=t$ and $b$)
\br
M_{\tilde{q}}^2 \ =\ \left ( \begin{array}{cc} M_{\tilde{q}LL}^2 & 
|M_{\tilde{q}LR}^2|~ e^{-i\phi_{\tilde{q}}} \\[2mm]
|M_{\tilde{q}RL}^2|~ e^{i\phi_{\tilde{q}}} & M_{\tilde{q}RR}^2 
\end{array} \right ) \;,
\er
is Hermitian and can be diagonalised by the unitary transformation
\br
U^\dagger_{\tilde{q}} M_{\tilde{q}}^2 U_{\tilde{q}} \ = \ 
{\rm diag} (M_{{\tilde{q}_1}}^2 , M_{{\tilde{q}_2}}^2) \;,
\er
with ($M_{{\tilde{q}_1}^2} < M_{{\tilde{q}_2}^2}$)
\br 
M_{\tilde{q}_{(1)[2]}}^2 \ =\ \frac{1}{2} \biggl \{
\left (M_{\tilde{q}LL}^2+ M_{\tilde{q}RR}^2 \right ) (-)[+]
\sqrt { \left (M_{\tilde{q}LL}^2 -  M_{\tilde{q}RR}^2 \right )^2 +
4~ |M_{\tilde{q}RL}|^4 } \biggr \}  
\er
and
\br\label{unitary}
U_{\tilde{q}} \ = \ \left ( \begin{array}{cc} \cos\theta_{\tilde{q}} &
-\sin\theta_{\tilde{q}}~ e^{i\phi_{\tilde{q}}} \\[2mm]
\sin\theta_{\tilde{q}}~ e^{-i\phi_{\tilde{q}}} & 
\cos\theta_{\tilde{q}} \end{array} \right ) \;,
\er
where $-\pi/2 \le \theta_{\tilde{q}} \le \pi/2$ and 
\br
\tan2\theta_{\tilde{q}} \ =\  \frac{2~|M_{\tilde{q}RL}^2|}{M_{\tilde{q}LL}^2
-  M_{\tilde{q}RR}^2 }\; ,
\er
\begin{equation}
\sin\phi_{\tilde{q}}=\frac{\Im (M_{\tilde{q}RL}^2)}
{|M_{\tilde{q}RL}^2|}\; ,
\end{equation}
where $\Im$ refers to the imaginary part of a complex quantity.

Now, in order to construct the Feynman rules for the Higgs-squark-squark
vertices, we need the transformation of the squark weak basis 
$(\tilde{q}_L, \tilde{q}_R)$ into 
 the mass basis $(\tilde{q}_1, \tilde{q}_2)$, namely
\br\label{convert}
\left ( \begin{array}{c} \tilde{q}_L \\[2mm] \tilde{q}_R \end{array} \right )
\ =\ U_{\tilde{q}} 
\left ( \begin{array}{c} \tilde{q}_1 \\[2mm] \tilde{q}_2 \end{array} \right )
\;.
\er
Furthermore, one has to proceed by also
transforming the Higgs boson weak basis
($H_1^0,H_2^0,H_2^+,H_2^-$) of the four complex fields into
the real eight physical ones ($H^0,h^0,A^0,H^+,H^-,G^0,G^+,G^-$). 
Following Ref.~\cite{HHG}, the transformation can be written as follows
\br
H_1^0 \ &=& v_1 + \frac{1}{\sqrt{2}} \left ( H^0 \cos\alpha -h^0 \sin\alpha
+i A^0\sin\beta - iG^0\cos\beta \right ) \;,\\[2mm]
H_2^0 \ &=& v_2 + \frac{1}{\sqrt{2}} \left ( H^0 \sin\alpha +h^0 \cos\alpha
+i A^0\cos\beta + i G^0 \sin\beta \right )\;,\\[2mm]
H_1^- \ &=& \ H^- \sin\beta - G^-\cos\beta \;,\\[2mm]
H_2^+ \ &=& \ H^+ \cos\beta + G^+\sin\beta \;,
\er
with $(H^+)^*\equiv H^-$ and 
\br
\sin2\alpha \ = \ -\sin2\beta \left (\frac{M_{H^0}^2 + M_{h^0}^2}
{M_{H^0}^2 - M_{h^0}^2} \right )\;,
\er
where $M_{H^0}, M_{h^0}$ are the tree-level CP-even Higgs boson masses.

The Feynman rules for the Higgs-squark-squark vertices,
involving mixing and phases, finally are (here, $\Phi=\Phi^0$ 
and $H^\pm$):
\begin{eqnarray}\label{mixing}
\lambda_{\Phi\tilde{q}_1\tilde{q}^{'*}_1} &=& c_{\sq} c_{\sq'} \lambda_{\Phi
\tilde{q}_L\tilde{q}^{'*}_L} + s_{\sq} s_{\sq'} e^{-i(\phi_{\sq}-
\phi_{\sq'})}
\lambda_{\Phi\tilde{q}_R \tilde{q}^{'*}_R}+
c_{\sq} s_{\sq'} e^{i \phi_{\tilde{q}'}}
\lambda_{\Phi\tilde{q}_L\tilde{q}^{'*}_R} + s_{\sq} c_{\sq'}
e^{-i \phi_{\tilde{q}}}
\lambda_{\Phi\tilde{q}_R\tilde{q}^{'*}_L} \;, \nonumber \\[2mm]
\lambda_{\Phi\tilde{q}_2\tilde{q}^{'*}_2} &=& s_{\sq} s_{\sq'}
e^{i(\phi_{\sq}-
\phi_{\sq'})} \lambda_{\Phi
\tilde{q}_L\tilde{q}^{'*}_L} + c_{\sq} c_{\sq'}
\lambda_{\Phi\tilde{q}_R\tilde{q}^{'*}_R}
-s_{\sq} c_{\sq'} e^{i \phi_{\tilde{q}}}
\lambda_{\Phi\tilde{q}_L\tilde{q}^{'*}_R}
-c_{\sq} s_{\sq'} e^{-i \phi_{\tilde{q}'}}
\lambda_{\Phi\tilde{q}_R\tilde{q}^{'*}_L} \; ,\nonumber \\[2mm]
\lambda_{\Phi\tilde{q}_1\tilde{q}^{'*}_2} &=&
-c_{\sq} s_{\sq'} e^{-i\phi_{\tilde{q}'}} \lambda_{\Phi
\tilde{q}_L\tilde{q}^{'*}_L} + s_{\sq} c_{\sq'} e^{-i \phi_{\tilde{q}}}
\lambda_{\Phi\tilde{q}_R\tilde{q}^{'*}_R}
+c_{\sq} c_{\sq'} \lambda_{\Phi\tilde{q}_L\tilde{q}^{'*}_R}
-s_{\sq} s_{\sq'} e^{-i (\phi_{\tilde{q}}+\phi_{\tilde{q}'})}
\lambda_{\Phi\tilde{q}_R\tilde{q}^{'*}_L} \; ,\nonumber \\[2mm]
\lambda_{\Phi\tilde{q}_2\tilde{q}^{'*}_1} &=&
-s_{\sq} c_{\sq'} e^{i\phi_{\tilde{q}}} \lambda_{\Phi
\tilde{q}_L\tilde{q}^{'*}_L} + c_{\sq} s_{\sq'} e^{i \phi_{\tilde{q}'}}
\lambda_{\Phi\tilde{q}_R\tilde{q}^{'*}_R}
-s_{\sq} s_{\sq'} e^{i(\phi_{\tilde{q}'}+\phi_{\tilde{q}})}
 \lambda_{\Phi\tilde{q}_L\tilde{q}^{'*}_R}
+c_{\sq} c_{\sq'}
\lambda_{\Phi\tilde{q}_R\tilde{q}^{'*}_L} \;. \label{fr}\nonumber  \\
\end{eqnarray}

For the case of stop squarks (i.e., ${\tilde q}={\tilde t}$), one has
\br
M_{\tilde{t}LL}^2 &=& M_{\tilde{Q}}^2+M_t^2+\frac{1}{6} (4 M_W^2-M_Z^2)\cos 
2\beta\; , \\
M_{\tilde{t}RR}^2 &=& M_{\tilde{U}}^2+M_t^2-\frac{2}{3} (M_W^2-M_Z^2)
\cos2\beta\; , \\
M_{\tilde{t}RL}^2 =( M_{\tilde{t}LR}^2)^* &=& m_t ( A_t + \mu^* \cot\beta )\; ,
\er
and thus
\br\label{tmix}
\tan 2\theta_{\tilde{t}} \ &=& \ \frac{2 m_t | A_t + \mu^* \cot\beta |}
{M_{\tilde{Q}}^2 - M_{\tilde{U}}^2 + \left ( \frac{4}{3}M_W^2 -
\frac{5}{6} M_Z^2 \right ) \cos 2\beta } \;, \label{thetat}\\[2mm]
\sin\phi_{\tilde{t}} \ &=& \ \frac{|A_t| \sin\phi_{A_t}-|\mu|\sin\phi_\mu
\cot\beta}{|A_t + \mu^* \cot\beta|} \label{ang}\;,
\er
where $\mu = |\mu|e^{i\phi_\mu}$ and $A_t=|A_t|e^{i\phi_{A_t}}$. 

Similarly, one obtains for sbottoms (i.e., ${\tilde q}={\tilde b}$):
\br
M_{\tilde{b}LL}^2 &=& M_{\tilde{Q}}^2+m_b^2-\frac{1}{6} (2 M_W^2+M_Z^2)\cos 
2\beta \; , \\
M_{\tilde{b}RR}^2 &=& M_{\tilde{D}}^2+m_b^2+\frac{1}{3} (M_W^2-M_Z^2)
\cos2\beta \; ,\\
M_{\tilde{b}RL}^2 =( M_{\tilde{b}LR}^2)^* &=& m_b ( A_b + \mu^* \tan\beta )\; ,
\er
with
\br\label{bmix}
\tan 2\theta_{\tilde{b}} \ &=& \ \frac{2 m_b | A_b + \mu^* \tan\beta |}
{M_{\tilde{Q}}^2 - M_{\tilde{D}}^2 + \left ( -\frac{2}{3}M_W^2 +
\frac{1}{6} M_Z^2 \right ) \cos 2\beta } \;, \\[2mm]
\sin\phi_{\tilde{b}} \ &=& \ \frac{|A_b| \sin\phi_{A_b}-|\mu|\sin\phi_\mu
\tan\beta}{|A_b + \mu^* \tan\beta|} \;,
\er
where  $A_b=|A_b|e^{i\phi_{A_b}}$.

The chiral couplings 
$\lambda_{\Phi\tilde{q}_{\chi}\tilde{q}_{\chi}^{'*}}$ ($\chi=L,R$) 
of eq.~(\ref{mixing}) can be found in Ref.~\cite{HHG}, with the
only exception of those cases where  $\mu$ and $A_{\tilde{q}}$ 
enter\footnote{In other terms,
for real $\mu$ and $A_{\tilde{q}}$ our expressions 
reduce to those  in Ref.~\cite{HHG}.}, for which one has to adopt the
following set of formulae (with $g^2\equiv 4\pi\alpha_{\rm{EM}}/\sin^2\theta_W$
the weak constant and where $M_W$ is the $W^\pm$ boson mass):
\br\label{LR}
\lambda_{A^0 \tilde{t}_L \tilde{t}_R^*} \ &=& \
-\frac{gm_u}{2M_W} \left ( \mu^* -A_u \cot\beta \right ) \;,\\[2mm]
\lambda_{A^0 \tilde{t}_L^* \tilde{t}_R} \ &=& \
-(\lambda_{A^0 \tilde{t}_L \tilde{t}_R^*})^* \;,\\[2mm]
\lambda_{A^0 \tilde{b}_L \tilde{b}_R^*} \ &=& \ 
-\frac{gm_d}{2M_W} \left ( \mu^* -A_d \tan\beta \right ) \;,\\[2mm]
\lambda_{A^0 \tilde{b}_L^* \tilde{b}_R} \ &=& \
-(\lambda_{A^0 \tilde{b}_L \tilde{b}_R^*})^* \;,\\[2mm]
\lambda_{H^0 \tilde{t}_L \tilde{t}_R^*} \ &=& \
-\frac{i g m_u}{2M_W\sin\beta} \left (\mu^* \cos\alpha + A_u \sin\alpha
\right ) \;,\\[2mm]
\lambda_{H^0 \tilde{t}_L^* \tilde{t}_R} \ &=& \
-(\lambda_{H^0 \tilde{t}_L \tilde{t}_R^*})^* \;,\\[2mm]
\lambda_{H^0 \tilde{b}_L \tilde{b}_R^*} \ &=& \
-\frac{i g m_d}{2 M_W \cos\beta} \left ( \mu^* \sin\alpha + A_d \cos\alpha
\right )\;,\\[2mm]
\lambda_{H^0 \tilde{b}_L^* \tilde{b}_R} \ &=& \
-(\lambda_{H^0 \tilde{b}_L \tilde{b}_R^*})^* \;,\\[2mm]
\lambda_{h^0 \tilde{t}_L \tilde{t}_R^*} \ &=& \
\frac{i g m_u}{2M_W\sin\beta} \left (\mu^* \sin\alpha - A_u \cos\alpha
\right ) \;,\\[2mm]
\lambda_{h^0 \tilde{t}_L^* \tilde{t}_R} \ &=& \
-(\lambda_{h^0 \tilde{t}_L \tilde{t}_R^*})^* \;,\\[2mm]
\lambda_{h^0 \tilde{b}_L \tilde{b}_R^*} \ &=& \
-\frac{i g m_d}{2 M_W \cos\beta} \left ( \mu^* \cos\alpha - A_d \sin\alpha
\right )\;,\\[2mm]\label{RL}
\lambda_{h^0 \tilde{b}_L^* \tilde{b}_R} \ &=& \
-(\lambda_{h^0 \tilde{b}_L \tilde{b}_R^*})^* \;,\\[2mm]
\lambda_{H^+ \tilde{b}_L \tilde{t}_R^*} \ &=& \
-\frac{i g m_u}{\sqrt{2} M_W} \left (\mu^* - A_u \cot\beta \right )\;,
 \\[2mm]
\lambda_{H^- \tilde{t}_L \tilde{b}_R^*} \ &=& \
-\frac{i g m_d}{\sqrt{2} M_W} \left (\mu^* - A_d \tan\beta \right )\;,
 \\[2mm]
\lambda_{H^+ \tilde{t}_L^* \tilde{b}_R} \ &=& \
-\frac{i g m_d}{\sqrt{2} M_W} \left (\mu - A_d^* \tan\beta \right )\;,
 \\[2mm]
\lambda_{H^- \tilde{b}_L^* \tilde{t}_R} \ &=& \
-\frac{i g m_u}{\sqrt{2} M_W} \left (\mu - A_u^* \cot\beta \right ) \;.
\er
(Although we have not made use of the Feynman rules involving charged
Higgses we display them here for completeness.) We need also the interactions
among Higgs bosons and quarks and these read as follows:
\br 
\lambda_{A^0t\bar{t}} \ &=& \ -\frac{g m_u \cot\beta}{2 M_W}\gamma_5\; ,\\[2mm]
\label{auu}
\lambda_{A^0b\bar{b}} \ &=& \ -\frac{g m_d \tan\beta}{2 M_W}\gamma_5\; ,\\[2mm]
\label{add}
\lambda_{H^0t\bar{t}} \ &=& \ -\frac{i g m_u \sin\alpha}{2 M_W \sin\beta}
\; ,\\[2mm]
\lambda_{H^0b\bar{b}} \ &=& \ -\frac{i g m_d \cos\alpha}{2 M_W \cos\beta}
\; ,\\[2mm]
\lambda_{h^0t\bar{t}} \ &=& \ -\frac{i g m_u \cos\alpha}{2 M_W \sin\beta}
\; ,\\[2mm]
\lambda_{h^0b\bar{b}} \ &=& \ \frac{i g m_d \sin\alpha}{2 M_W \cos\beta}
\; .
\er

It is now useful to look at the explicit phase dependence of 
the vertices involving Higgs bosons and squarks: to this end,
 we expand
our formulae (\ref{fr}).  The relevant couplings involving
 the CP-odd Higgs boson read as\footnote{Note that the
vertices with gluons and squarks are not affected by the phases.}:
\br
\lambda_{A^0\tilde{t}_1\tilde{t}_1^*}
 &=& -i\biggl [\frac{g m_t }{2 M_W} \biggr ]
\sin2\theta_{\tilde{t}} \biggl \{|\mu|\sin(\phi_{\tilde{t}}-\phi_\mu)
-|A_t|\sin(\phi_{\tilde{t}}+\phi_{A_t})\cot\beta \biggr \} \;,
 \label{At1t1}\nonumber \\[2mm]
\lambda_{A^0\tilde{t}_2\tilde{t}_2^*} &=& 
-\lambda_{A^0\tilde{t}_1\tilde{t}_1^*}\;,
 \nonumber \\[2mm]
\lambda_{A^0\tilde{b}_1\tilde{b}_1^*}
 &=& -i\biggl [\frac{g m_b }{2 M_W} \biggr ]
\sin2\theta_{\tilde{b}} \biggl \{|\mu|\sin(\phi_{\tilde{b}}-\phi_\mu)
-|A_b|\sin(\phi_{\tilde{b}}+\phi_{A_b})\tan\beta \biggr \} \;, 
\nonumber \\[2mm]
\lambda_{A^0\tilde{b}_2\tilde{b}_2^*} &=& -
\lambda_{A^0\tilde{b}_1\tilde{b}_1^*} \;.
\label{Aqq}
\er
For the CP-even Higgs bosons we find (here, $s_W\equiv\sw$),
\br
\lambda_{H^0\tilde{t}_1\tilde{t}_1^*} &=& \Biggl [\frac{i g M_Z}{c_W}\Biggr ]
\biggl \{-\biggl [\frac{1}{2}\cos^2\theta_{\tilde{t}}-e_u s_W^2 
\cos2\theta_{\tilde{t}} \biggr ]\cos(\alpha+\beta)-\frac{m_t^2}{M_Z^2}
\frac{\sin\alpha}{\sin\beta} \label{Ht1t1}
\nonumber \\[2mm]
&-& \frac{m_t \sin2\theta_{\tilde{t}}}{2 M_Z^2 \sin\beta} \biggl [
|\mu|\cos(\phi_{\tilde{t}}-\phi_\mu)\cos\alpha
+|A_t|\cos(\phi_{\tilde{t}}+\phi_{A_t})
\sin\alpha \biggr ] \Biggr \} \;, \\[4mm]
\lambda_{H^0\tilde{t}_2\tilde{t}_2^*} &=& \Biggl [\frac{i g M_Z}{c_W}\Biggr ]
\biggl \{-\biggl [\frac{1}{2}\sin^2\theta_{\tilde{t}}+e_u s_W^2 
\cos2\theta_{\tilde{t}} \biggr ]\cos(\alpha+\beta)-\frac{m_t^2}{M_Z^2}
\frac{\sin\alpha}{\sin\beta}
\nonumber \\[2mm]
&+& \frac{m_t \sin2\theta_{\tilde{t}}}{2 M_Z^2 \sin\beta} \biggl [
|\mu|\cos(\phi_{\tilde{t}}-\phi_\mu)\cos\alpha
+|A_t|\cos(\phi_{\tilde{t}}+\phi_{A_t})
\sin\alpha \biggr ] \Biggr \} \;, \\[4mm]
\lambda_{H^0\tilde{b}_1\tilde{b}_1^*} &=& \Biggl [\frac{i g M_Z}{c_W}\Biggr ]
\biggl \{\biggl [\frac{1}{2}\cos^2\theta_{\tilde{b}}+e_d s_W^2 
\cos2\theta_{\tilde{b}} \biggr ]\cos(\alpha+\beta)-\frac{m_b^2}{M_Z^2}
\frac{\cos\alpha}{\cos\beta}
\nonumber \\[2mm]
&-& \frac{m_b \sin2\theta_{\tilde{b}}}{2 M_Z^2 \cos\beta} \biggl [
|\mu|\cos(\phi_{\tilde{b}}-\phi_\mu)\sin\alpha
+|A_b|\cos(\phi_{\tilde{b}}+\phi_{A_b})
\cos\alpha \biggr ] \Biggr \} \;, \\[4mm]
\lambda_{H^0\tilde{b}_2\tilde{b}_2^*} &=& \Biggl [\frac{i g M_Z}{c_W}\Biggr ]
\biggl \{\biggl [\frac{1}{2}\sin^2\theta_{\tilde{b}}-e_d s_W^2 
\cos2\theta_{\tilde{b}} \biggr ]\cos(\alpha+\beta)-\frac{m_b^2}{M_Z^2}
\frac{\cos\alpha}{\cos\beta}
\nonumber \\[2mm]
&+& \frac{m_b \sin2\theta_{\tilde{b}}}{2 M_Z^2 \cos\beta} \biggl [
|\mu|\cos(\phi_{\tilde{b}}-\phi_\mu)\sin\alpha
+|A_b|\cos(\phi_{\tilde{b}}+\phi_{A_b})
\cos\alpha \biggr ] \Biggr \} \;, \\[4mm]
\lambda_{h^0\tilde{t}_1\tilde{t}_1^*} &=& \Biggl [\frac{i g M_Z}{c_W}\Biggr ]
\biggl \{\biggl [\frac{1}{2}\cos^2\theta_{\tilde{t}}-e_u s_W^2 
\cos2\theta_{\tilde{t}} \biggr ]\sin(\alpha+\beta)-\frac{m_t^2}{M_Z^2}
\frac{\cos\alpha}{\sin\beta}
\label{ht1t1}\nonumber \\[2mm]
&+& \frac{m_t \sin2\theta_{\tilde{t}}}{2 M_Z^2 \sin\beta} \biggl [
|\mu|\cos(\phi_{\tilde{t}}-\phi_\mu)\sin\alpha
-|A_t|\cos(\phi_{\tilde{t}}+\phi_{A_t})
\cos\alpha \biggr ] \Biggr \} \;, \\[4mm]
\lambda_{h^0\tilde{t}_2\tilde{t}_2^*} &=& \Biggl [\frac{i g M_Z}{c_W}\Biggr ]
\biggl \{\biggl [\frac{1}{2}\sin^2\theta_{\tilde{t}}+e_u s_W^2 
\cos2\theta_{\tilde{t}} \biggr ]\sin(\alpha+\beta)-\frac{m_t^2}{M_Z^2}
\frac{\cos\alpha}{\sin\beta}
\nonumber \\[2mm]
&-& \frac{m_t \sin2\theta_{\tilde{t}}}{2 M_Z^2 \sin\beta} \biggl [
|\mu|\cos(\phi_{\tilde{t}}-\phi_\mu)\sin\alpha
-|A_t|\cos(\phi_{\tilde{t}}+\phi_{A_t})
\cos\alpha \biggr ] \Biggr \} \;, \\[4mm]
\lambda_{h^0\tilde{b}_1\tilde{b}_1^*} &=& \Biggl [\frac{i g M_Z}{c_W}\Biggr ]
\biggl \{-\biggl [\frac{1}{2}\cos^2\theta_{\tilde{b}}+e_d s_W^2 
\cos2\theta_{\tilde{b}} \biggr ]\sin(\alpha+\beta)+\frac{m_b^2}{M_Z^2}
\frac{\sin\alpha}{\cos\beta}
\nonumber \\[2mm]
&-& \frac{m_b \sin2\theta_{\tilde{b}}}{2 M_Z^2 \cos\beta} \biggl [
|\mu|\cos(\phi_{\tilde{b}}-\phi_\mu)\cos\alpha
-|A_b|\cos(\phi_{\tilde{b}}+\phi_{A_b})
\sin\alpha \biggr ] \Biggr \} \;, \\[4mm]
\lambda_{h^0\tilde{b}_2\tilde{b}_2^*} &=& \Biggl [\frac{i g M_Z}{c_W}\Biggr ]
\biggl \{-\biggl [\frac{1}{2}\sin^2\theta_{\tilde{b}}-e_d s_W^2 
\cos2\theta_{\tilde{b}} \biggr ]\sin(\alpha+\beta)+\frac{m_b^2}{M_Z^2}
\frac{\sin\alpha}{\cos\beta}
\nonumber \\[2mm]
&+& \frac{m_b \sin2\theta_{\tilde{b}}}{2 M_Z^2 \cos\beta} \biggl [
|\mu|\cos(\phi_{\tilde{b}}-\phi_\mu)\cos\alpha
-|A_b|\cos(\phi_{\tilde{b}}+\phi_{A_b})
\sin\alpha \biggr ] \Biggr \} \;, 
\label{hb2b2}
\er
where $e_u=+2/3$ and $e_d=-1/3$. Of course, one can get the corresponding
vertices for all squarks flavours by a simple substitution $t\ar u,c$ and 
$b\ar d,s$.

\nnappend{Appendix B: matrix elements and cross sections}

Since, as we have already discussed in the main body of the paper,
the CP-violating phases induce a non-zero contribution from squark
loops in CP-odd Higgs boson production which is absent in the
phaseless MSSM and since this has not been 
evaluated yet in the literature, we had to perform the loop tensor reduction 
in such a case from scratch. However, for comparison purposes, we have 
also recalculated the well known tensor associated to CP-even Higgs boson 
production and found agreement with old results.
We have evaluated the diagrams in the  
(modified) Dimensional Regularisation ($\overline{DR}$) scheme, which
preserves the SUSY Ward identities of the theory
up to two loops. This enabled us to
check analytically the gauge invariance of our results. Furthermore, we
have carried out  the $\gamma$-algebra in four dimensions while using
analytical continuation in $d$ dimensions in order
to calculate the  divergent parts of the  integrals. 
The squared matrix elements summed/averaged over final/initial spins
and colours finally
are\footnote{Note that the $\gamma_5$-matrix  is here intended
to be removed from the expressions of the $\lambda_{A^0q\bar q}$'s of 
eqs.~(\ref{auu})--(\ref{add}).}: 
\br
|\overline{\cal M}|^2_{gg\rightarrow h^0} &=& 
\frac{\alpha^2_s(Q) M_{h^0}^4}{256 \pi^2}
\Biggl |\sum_{q}\frac{\lambda_{h^0q\bar q}}{m_q}\tau_q\biggl [ 1+(1-\tau_q)
f(\tau_q) \biggr ] -\frac{1}{4} \sum_{\tilde{q}}\frac{\lambda_{h^0\tilde{q}
\tilde{q}^*}}
{m_{\tilde{q}}^2} \tau_{\tilde{q}} \biggl [
1-\tau_{\tilde{q}} f(\tau_{\tilde{q}}) \biggr ] \Biggr |^2 \;,\nonumber \\[2mm]
|\overline{\cal M}|^2_{gg\rightarrow H^0} &=& 
\frac{\alpha^2_s(Q) M_{H^0}^4}{256 \pi^2}
\Biggl |\sum_{q}\frac{\lambda_{H^0q\bar q}}{m_q}\tau_q\biggl [ 1+(1-\tau_q)
f(\tau_q) \biggr ] -\frac{1}{4} \sum_{\tilde{q}}\frac{\lambda_{H^0\tilde{q}
\tilde{q}^*}}
{m_{\tilde{q}}^2} \tau_{\tilde{q}} \biggl [
1-\tau_{\tilde{q}} f(\tau_{\tilde{q}}) \biggr ] \Biggr |^2 \;,\nonumber \\[2mm]
|\overline{\cal M}|^2_{gg\rightarrow A^0}&=&
\frac{\alpha^2_s(Q) M_{A^0}^4}{256 \pi^2}
\Biggl \{ \Biggl | \sum_q \frac{\lambda_{A^0q\bar q}}{m_q} 
\biggl [\tau_q f(\tau_q)
\biggr ]\biggr |^2 +\frac{1}{16}\biggl | \sum_{\tilde{q}}
\frac{\lambda_{A^0\tilde{q}\tilde{q}^*}}{m_{\tilde{q}}^2} \biggl [
\tau_{\tilde{q}} \biggl ( 1-\tau_{\tilde{q}} f(\tau_{\tilde{q}}) \biggr ) 
\biggr ] \Biggr |^2 \Biggr \} 
\label{me}
\er
where $\tau_{q,\tilde{q}}=\frac{4 m_{q,\tilde{q}}^2}{M_{\Phi^0}^2}$, 
$q=t,b$ and 
$\tilde{q}=\tilde{t}_1,\tilde{t}_2,\tilde{b}_1,\tilde{b}_2$.
The function $f(\tau)$ stands for
\br
f(\tau) \ =\ -\frac{1}{2}\int_0^1 \frac{dy}{y}\ln\biggl 
(1-\frac{4 y (1-y)}{\tau} \biggr ) = \Biggl \{ \begin{array}{c}
~{\rm arcsin}^2(\frac{1}{\sqrt{\tau}}), ~~~~~~~~~~~~~~~\tau \ge 1\; , 
\\[2mm]
-\frac{1}{4} \biggl [ \ln\frac{1+\sqrt{1-\tau}}{1-\sqrt{1-\tau}} -
i\pi \biggr ]^2, ~~~~\tau < 1 \;,\end{array}
\er
and some useful limits are
\br
\lim_{{\tau}\rightarrow \infty}\biggl \{\tau \biggl [1+(1-\tau)f(\tau)\biggr ]
\biggr \}=+\frac{2}{3} \;,\\
\lim_{{\tau}\rightarrow \infty}\biggl \{\tau \biggl [1-\tau f(\tau) \biggr ]
\biggr \}=-\frac{1}{3} \; ,\\
\lim_{{\tau}\rightarrow \infty}\biggl \{\tau f(\tau)\biggr \}= +1\;.
\er
The non-existence of interference terms between quark and squark
loops for CP-odd Higgs boson production can readily be understood 
by looking at the corresponding amplitude formula ($P_1$, $P_2$
are the gluon four-momenta, $\epsilon_\mu(P_1)$, $\epsilon_\nu(P_2)$ their
polarisation four-vectors  and $a,b$ their colours)
\br
& &i\epsilon_\mu(P_1)\epsilon_\nu(P_2)
{\cal M}^{\mu\nu}_{ab}(gg\rightarrow A^0)=
-\frac{\alpha_s(Q)}{2\pi}\delta_{ab}\, \epsilon_\mu(P_1)\epsilon_\nu(P_2)
 \, \times \nonumber  \\[2mm]
& & \Biggl \{i \varepsilon^{\mu\nu\rho\sigma}P_{1\rho}P_{2\sigma}\sum_q
\frac{\lambda_{A^0q\bar q}}{m_q}\biggl [\tau_q f(\tau_q) \biggl ] +
\frac{1}{4}\sum_{\tilde{q}}\frac{\lambda_{A^0\tilde{q}\tilde{q}^*}}
{m_{\tilde{q}}^2} \biggl (g^{\mu\nu}P_1\cdot P_2-P_1^\nu P_2^\mu\biggr )
\tau_{\tilde{q}}\biggl [1-\tau_{\tilde{q}}f(\tau_{\tilde{q}})\biggr ]\Biggr \}
\nonumber  \;, \\ 
\label{oddme}
\er
where one notices an antisymmetric part -- note the
Levi-Civita tensor $\varepsilon$ -- associated to the quark contributions
(first term on the right-hand side) and a
symmetric one associated to the squark loops (second term 
on the right-hand side). For completeness, we
also give the tensor structure of the loop amplitudes corresponding to
 CP-even Higgs boson production:
\br
& &i\epsilon_\mu(P_1)\epsilon_\nu(P_2)
{\cal M}^{\mu\nu}_{ab}(gg\rightarrow h^0,H^0)=
\frac{\alpha_s(Q)}{2\pi}\delta_{ab}\epsilon_\mu(P_1)\epsilon_\nu(P_2)
\Biggl ( g^{\mu\nu}P_1\cdot P_2-P_1^\nu P_2^\mu\biggr )~\times \nonumber
 \\[2mm]
& &\biggl \{ \sum_q \frac{\lambda_{(h^0,H^0)q\bar q}}{m_q}\tau_q \biggl [
1+ (1-\tau_q)f(\tau_q)\biggr ] -\frac{1}{4}\sum_{\tilde{q}}
\frac{\lambda_{(h^0,H^0)\tilde{q}\tilde{q}^*}}{m^2_{\tilde{q}}}\tau_{\tilde{q}}
\biggl [1-\tau_{\tilde{q}}f(\tau_{\tilde{q}}) \biggr ] \Biggr \} \;.
\nonumber \\ 
\label{evenme}
\er
Here, interference effects clearly exist between the SM- and SUSY-like 
parts, because of  the symmetric nature
of both contributions.

The LO partonic cross sections at the energy
$\sqrt{\hat{s}}$ are then
\br
{\hat{\sigma}}^{\Phi^0}_{LO} \ =\ \frac{\pi}{\hat{s}}~
 |\overline{\cal M}|^2_{gg\to \Phi^0}
~\delta({\hat{s}}-M_{\Phi^0}^2) \;,
\er
whereas the corresponding 
hadronic rates for a collider CM energy  $\sqrt s$ read as
\br
\sigma^{\Phi^0}_{LO} \ =\ \frac{\pi}{M_{\Phi^0}^4}~ 
|\overline{\cal M}|^2_{gg\to \Phi^0} ~\tau
~\frac{d{\cal L}}{d\tau} \;,
\er
with $\tau=\frac{\hat{s}}{s}\equiv\frac{M_{\Phi^0}^2}{s}$ and
\br
\tau \frac{d{\cal L}}{d\tau} = \int_\tau^1 \frac{d x}{x} {\tau} g(x,Q)
 g\bigl (\frac{\tau}{x},Q \bigr ) \;,
\er
where $g(x,Q)$ is the PDF of the gluon, evaluated at the scale
$Q\equiv M_{\Phi^0}$.

As already mentioned in the Introduction, in performing
the two-loop analysis, we have made use of the formulae
given in Ref.~\cite{DDS}. We refer the reader to that paper
for specific details.

\newpage

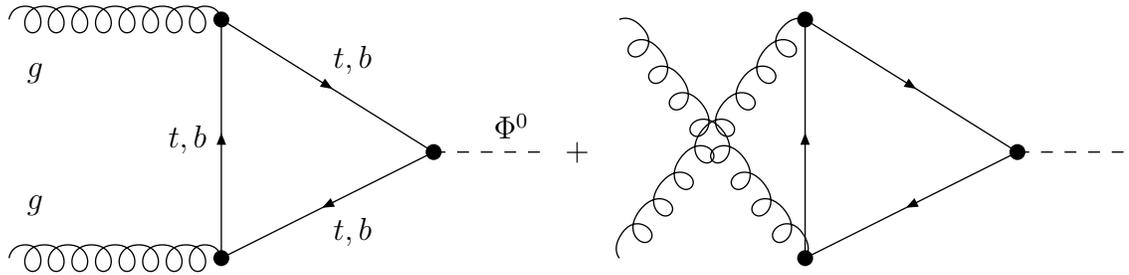
\begin{figure}
\begin{center}
\begin{picture}(0,0)(200,100)
\Gluon(20,0)(100,0){5}{8}\Vertex(100,0){3}
\Gluon(20,90)(100,90){5}{8}\Vertex(100,90){3}
\ArrowLine(100,0)(100,90)
\ArrowLine(100,90)(180,40)\Vertex(180,40){3}
\ArrowLine(180,40)(100,0)
\DashLine(180,40)(220,40){5}
\Text(30,20)[]{$g$}
\Text(30,70)[]{$g$}
\Text(210,50)[]{$\Phi^0$}
\Text(150,75)[]{$t,b$}
\Text(150,10)[]{$t,b$}
\Text(87.5,45)[]{$t,b$}
\Text(235,40)[]{+}
\Gluon(250,0)(320,90){5}{8}\Vertex(320,0){3}
\Gluon(250,90)(320,0){5}{8}\Vertex(320,90){3}
\ArrowLine(320,0)(320,90)
\ArrowLine(320,90)(400,40)\Vertex(400,40){3}
\ArrowLine(400,40)(320,0)
\DashLine(400,40)(440,40){5}
\end{picture}\\[5cm]
\caption{SM-like  contributions from top ($t$) and bottom ($b$) quarks to 
Higgs boson production via $gg\to \Phi^0$ in the MSSM.}
\label{fig:smdiag}
\end{center}
\end{figure}
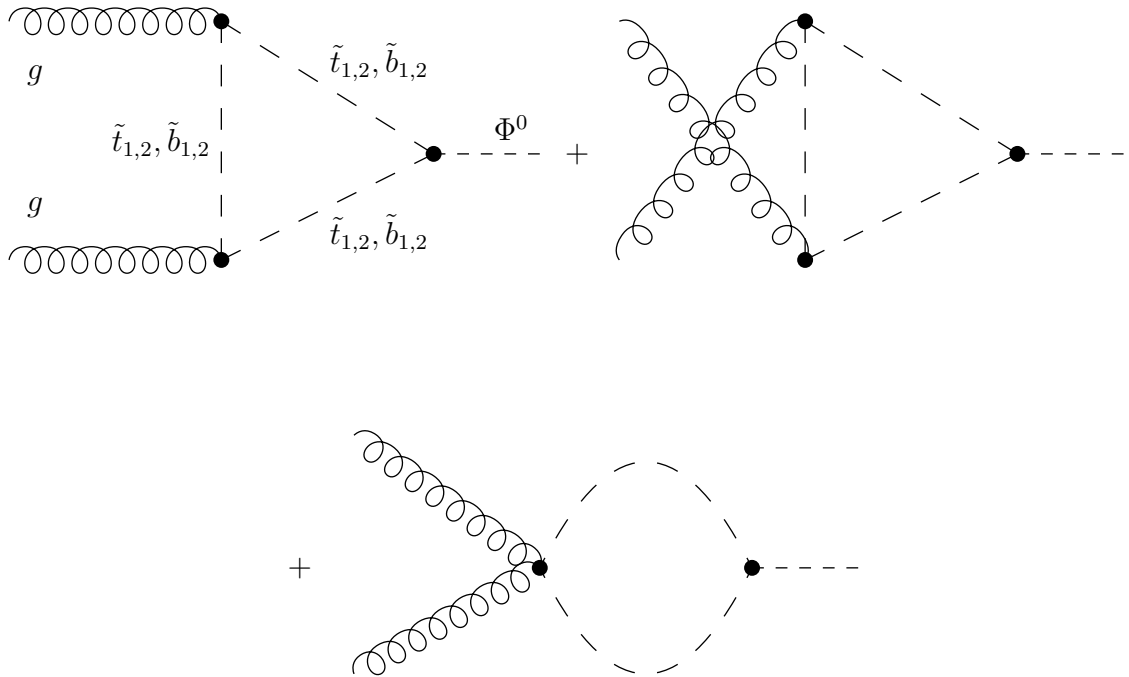
\begin{figure}
\begin{center}
\begin{picture}(0,0)(200,100)
\Gluon(20,0)(100,0){5}{8}\Vertex(100,0){3}
\Gluon(20,90)(100,90){5}{8}\Vertex(100,90){3}
\DashLine(100,0)(100,90){8}
\DashLine(100,90)(180,40){8}\Vertex(180,40){3}
\DashLine(180,40)(100,0){8}
\DashLine(180,40)(220,40){5}
\Text(30,20)[]{$g$}
\Text(30,70)[]{$g$}
\Text(210,50)[]{$\Phi^0$}
\Text(160,75)[]{$\tilde{t}_{1,2},\tilde{b}_{1,2}$}
\Text(160,10)[]{$\tilde{t}_{1,2},\tilde{b}_{1,2}$}
\Text(77.5,45)[]{$\tilde{t}_{1,2},\tilde{b}_{1,2}$}
\Text(235,40)[]{+}
\Gluon(250,0)(320,90){5}{8}\Vertex(320,0){3}
\Gluon(250,90)(320,0){5}{8}\Vertex(320,90){3}
\DashLine(320,0)(320,90){8}
\DashLine(320,90)(400,40){8}\Vertex(400,40){3}
\DashLine(400,40)(320,0){8}
\DashLine(400,40)(440,40){5}
\end{picture}\\[5cm] 
\end{center}
\begin{center}
\begin{picture}(0,0)(200,100)
\Text(130,40)[]{+}
\Gluon(150,0)(220,40){5}{8}\Vertex(220,40){3}
\Gluon(150,90)(220,40){5}{8}
\DashCurve{(220,40)(260,80)(300,40)}{8}
\DashCurve{(220,40)(260,0)(300,40)}{8}\Vertex(300,40){3}
\DashLine(300,40)(340,40){5}
\end{picture}\\[5cm] 
\caption{SUSY-like  contributions from top ($\tilde{t}_{1,2}$) 
and bottom ($\tilde{b}_{1,2}$) squarks to 
Higgs boson production via $gg\to \Phi^0$ in the MSSM. 
(Notice that, if the CP-symmetry is conserved, then $\Phi^0 \ne A^0$.)}
\label{fig:mssmdiag}
\end{center}
\end{figure}

\begin{figure}
\centerline{\hbox{\epsfig{figure=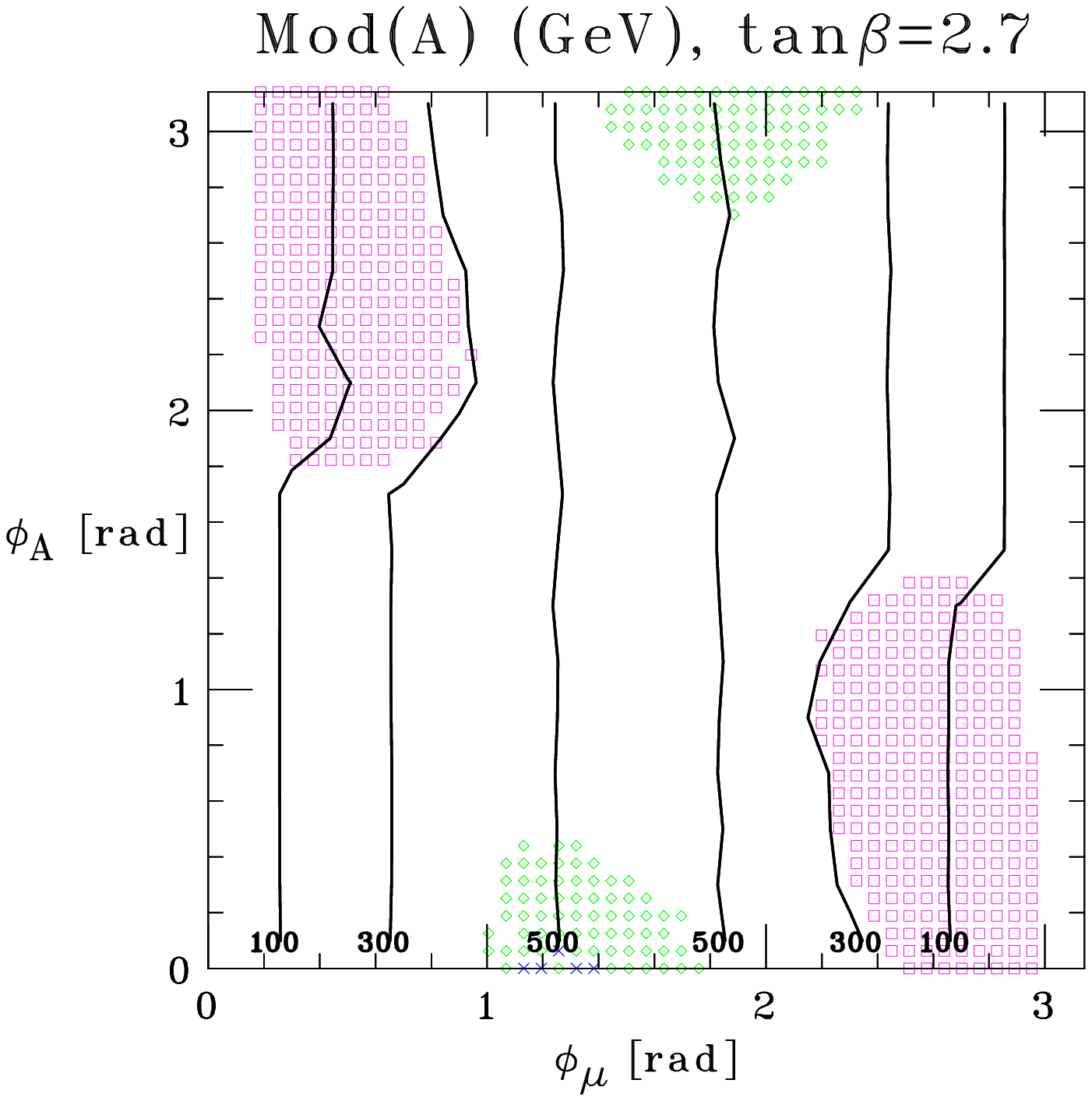,angle=90,height=3.25in}
\epsfig{figure=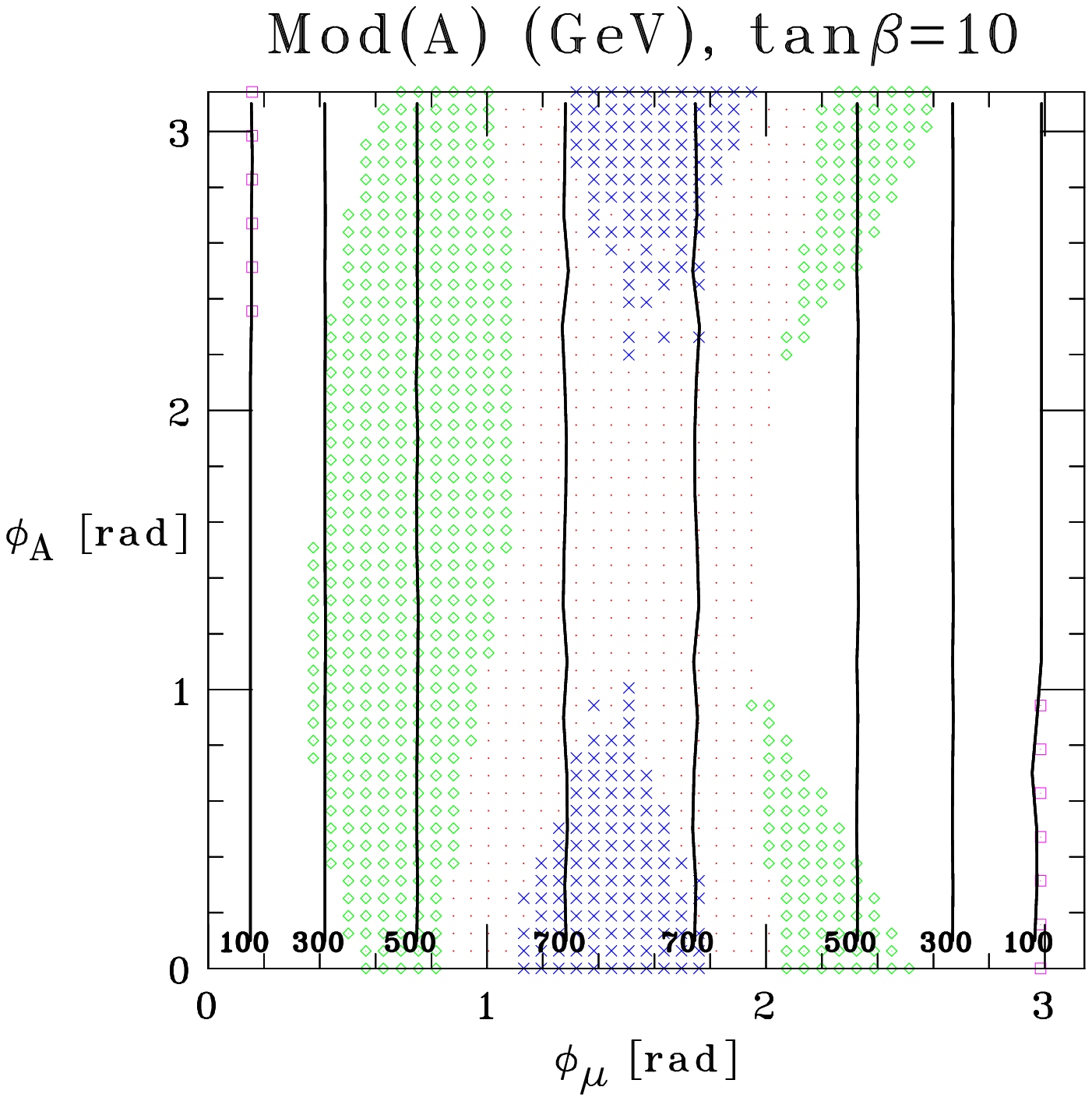,angle=90,height=3.25in}}}
\caption{Contour plots for the
values of the modulus of the common trilinear 
coupling, $|A|$, needed in order to obtain the cancellations of the 
SUSY contributions to the one-loop 
EDMs, over the ($\phi_\mu,\phi_A$) plane for small (left-hand plot) and 
large (right-hand plot) $\tan\beta$. The other MSSM parameters are as given in
Tab. I. Here and in the following, 
``$\times$''      symbols denote points excluded because of the 
negativity of the squark masses squared;
``$\diamond\hskip-0.160cm{\cdot}$'' symbols denote points excluded 
                  by the two-loop Zee-Barr
type contributions to the EDMs;  
``${\squaresm}{\hskip-0.220cm{\cdot}}$ '' 
and ``$\cdot$'' symbols denote 
points excluded from
Higgs boson and squark direct searches, respectively.}
\label{fig:A}
\end{figure}

\begin{figure}
\centerline{\hbox{\epsfig{figure=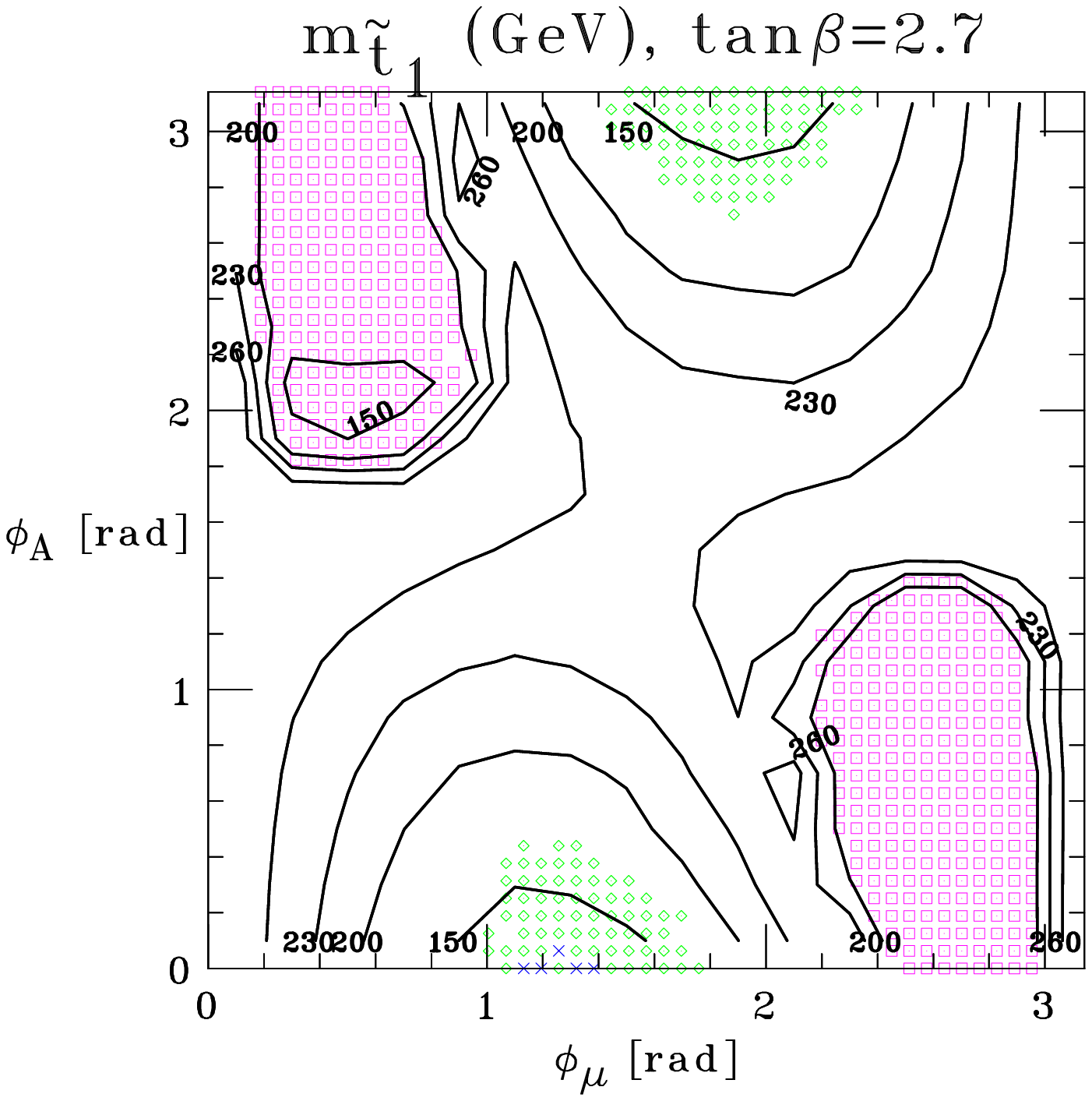,angle=90,height=3.25in}
\epsfig{figure=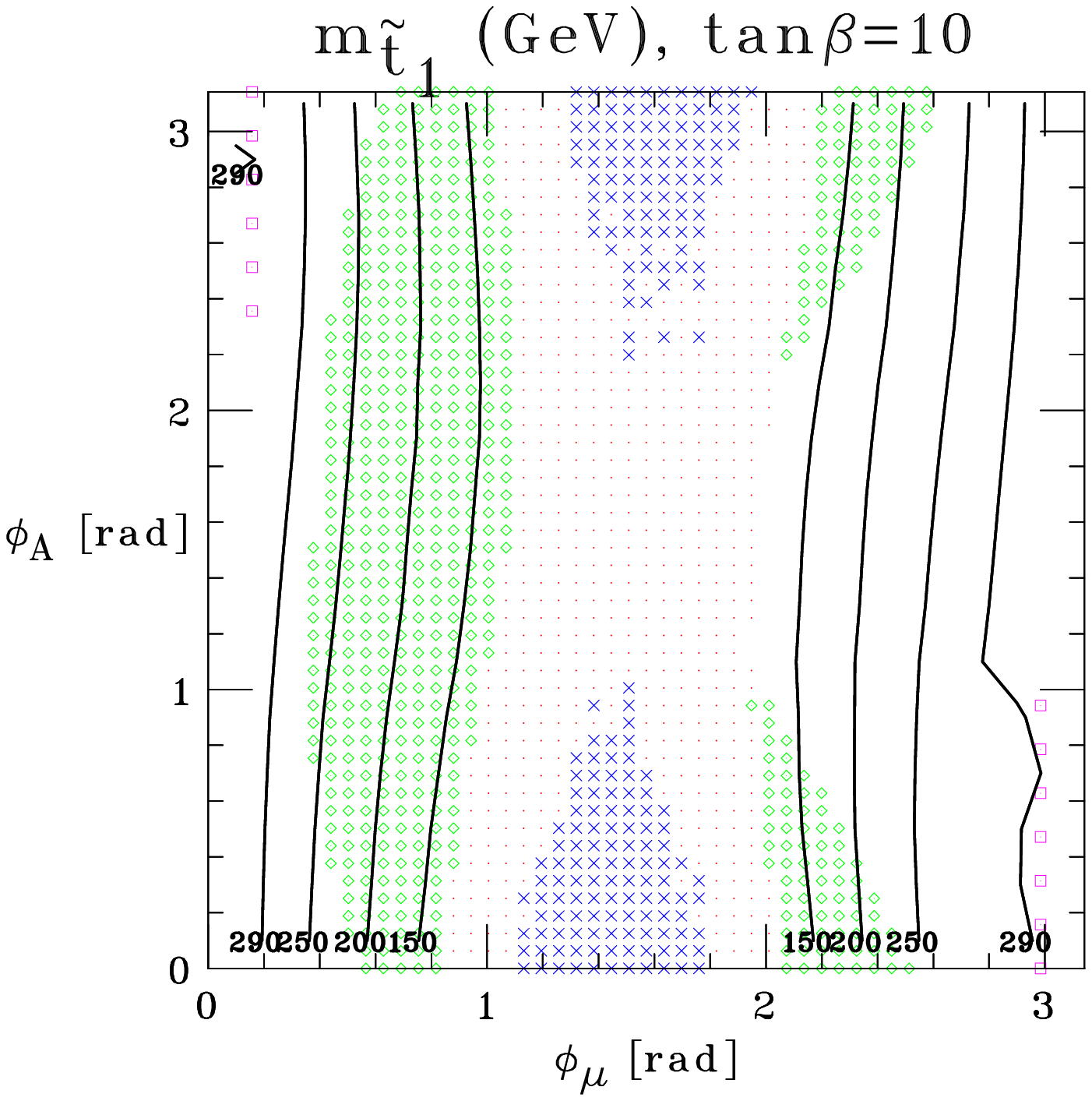,angle=90,height=3.25in}}}
\caption{Contour plots for the
values of the lightest top squark mass, $m_{{\tilde{t}}_1}$, corresponding
to those of  $|A|$ in Fig.~\ref{fig:A}, over the 
($\phi_\mu,\phi_A$) plane for small (left-hand plot) and 
large (right-hand plot) $\tan\beta$. The other MSSM parameters are as given in
Tab. I.}
\label{fig:stop1}
\end{figure}

\begin{figure}
\centerline{\hbox{\epsfig{figure=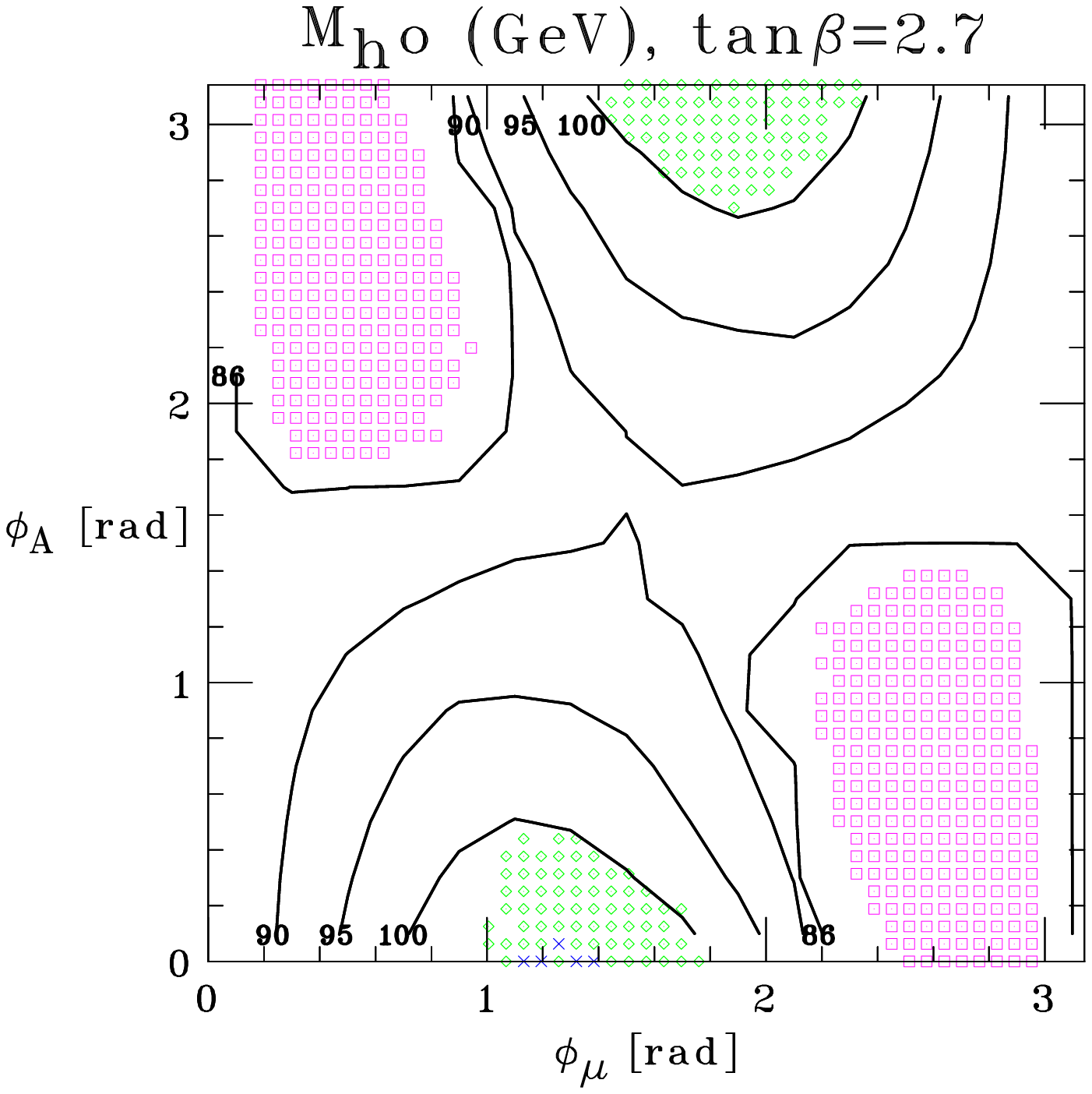,angle=90,height=3.25in}
\epsfig{figure=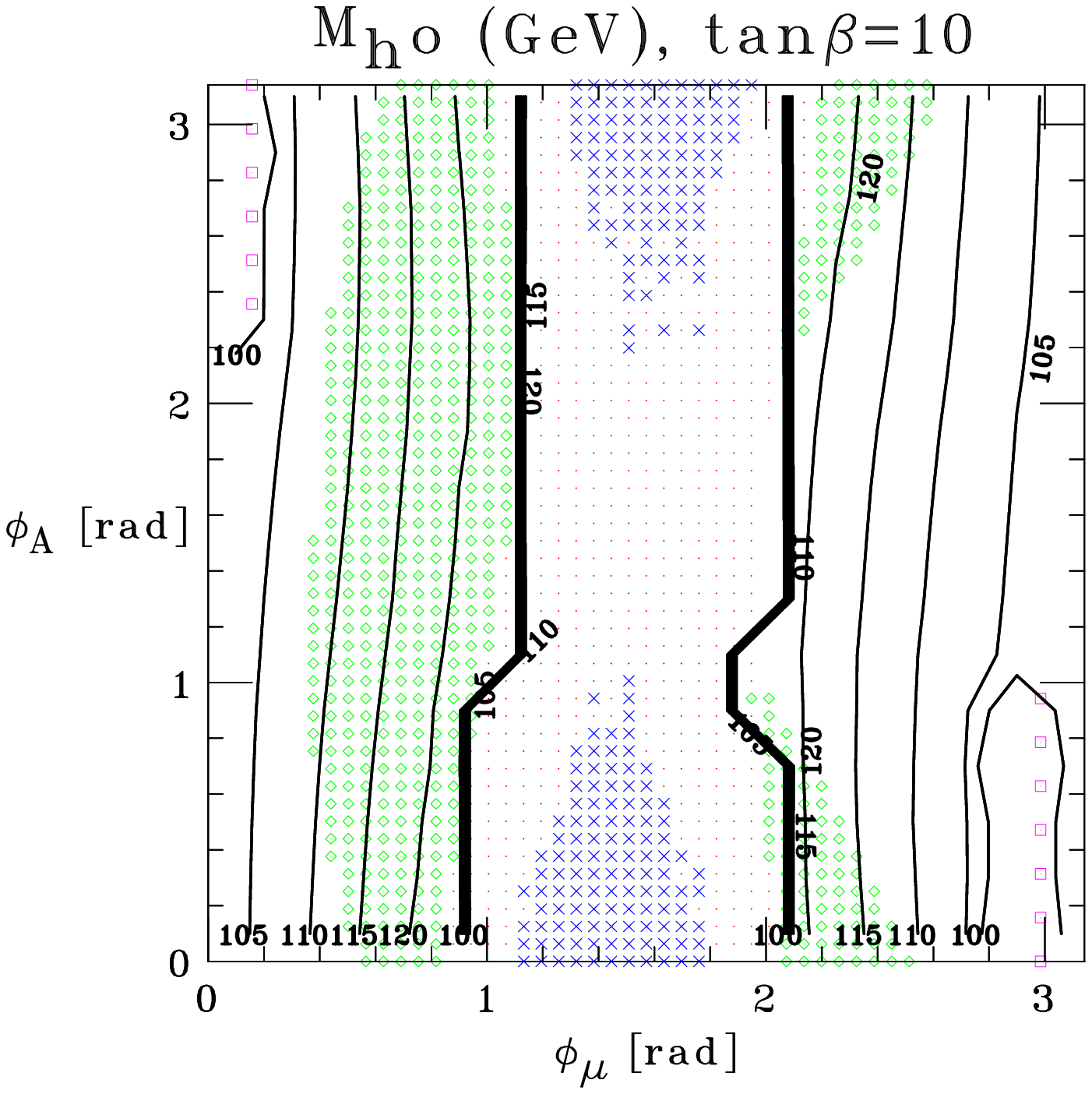,angle=90,height=3.25in}}}
\caption{Contour plots for the
values of the lightest Higgs boson mass, $M_{h^0}$, corresponding
to those  of $|A|$ in Fig.~\ref{fig:A}, over the 
($\phi_\mu,\phi_A$) plane for small (left-hand plot) and 
large (right-hand plot) $\tan\beta$. The other MSSM parameters are as given in
Tab. I.}
\label{fig:higgs}
\end{figure}

\begin{figure}
\centerline{\hbox{\epsfig{figure=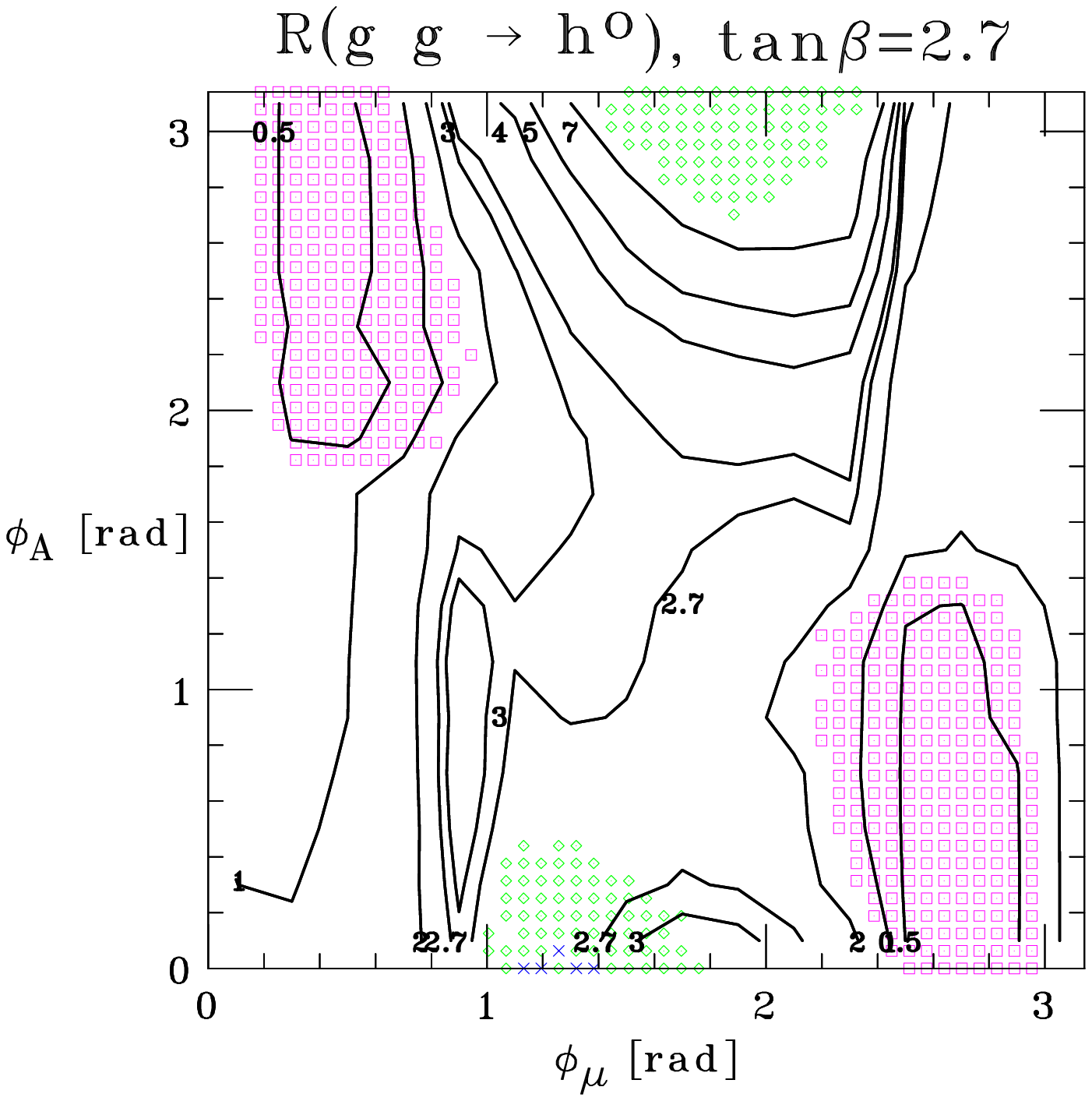,angle=90,height=3.25in}
\epsfig{figure=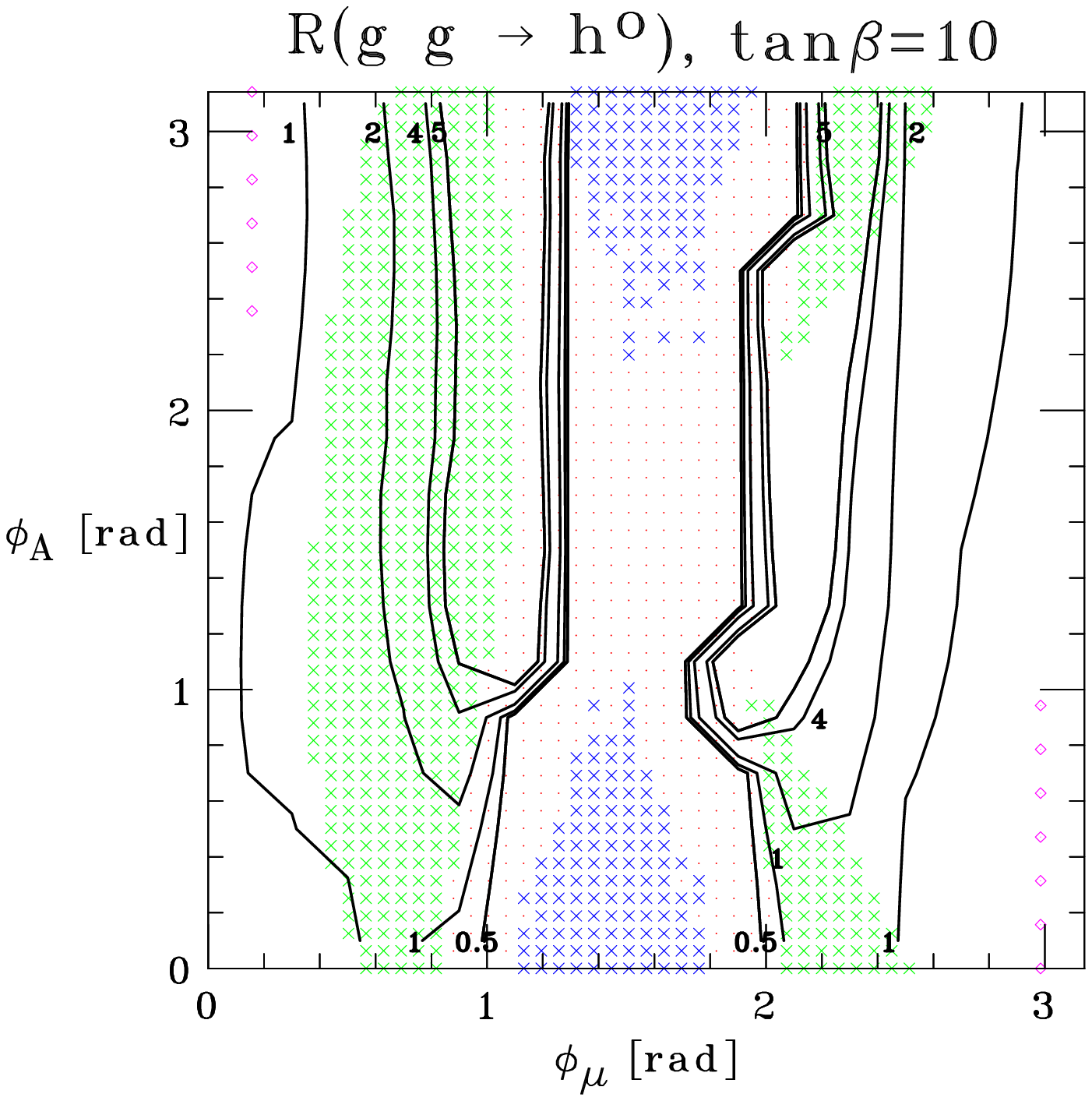,angle=90,height=3.25in}}}
\caption{Contour plots for the values of the ratio
in eq.~(\ref{rat}) for the case $\Phi^0=h^0$,
corresponding
to those  of $|A|$ in Fig.~\ref{fig:A}, over the 
($\phi_\mu,\phi_A$) plane for small (left-hand plot) and 
large (right-hand plot) $\tan\beta$. The other MSSM parameters are as given in
Tab. I.}
\label{fig:h0}
\end{figure}

\begin{figure}
\centerline{\hbox{\epsfig{figure=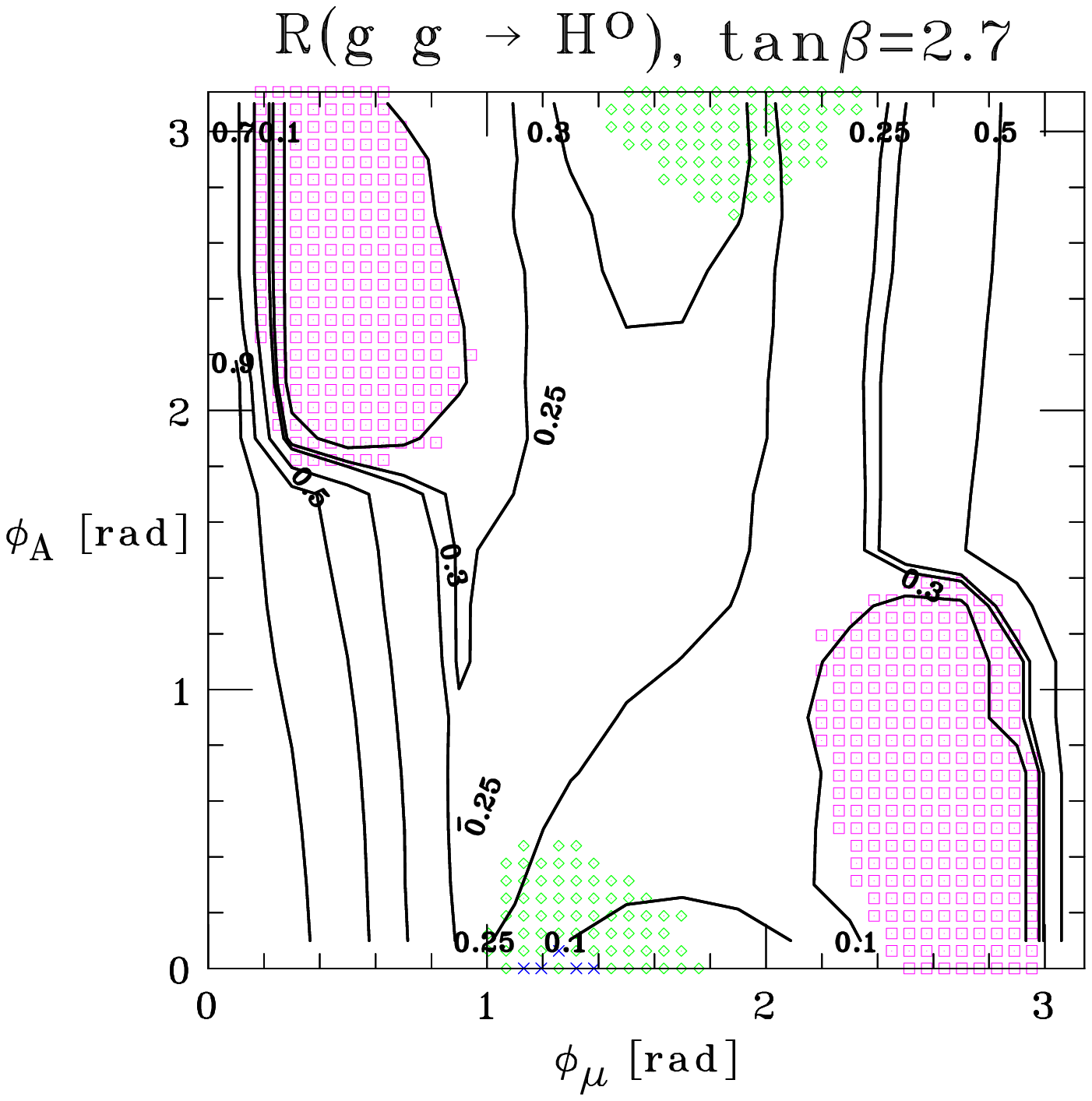,angle=90,height=3.25in}
\epsfig{figure=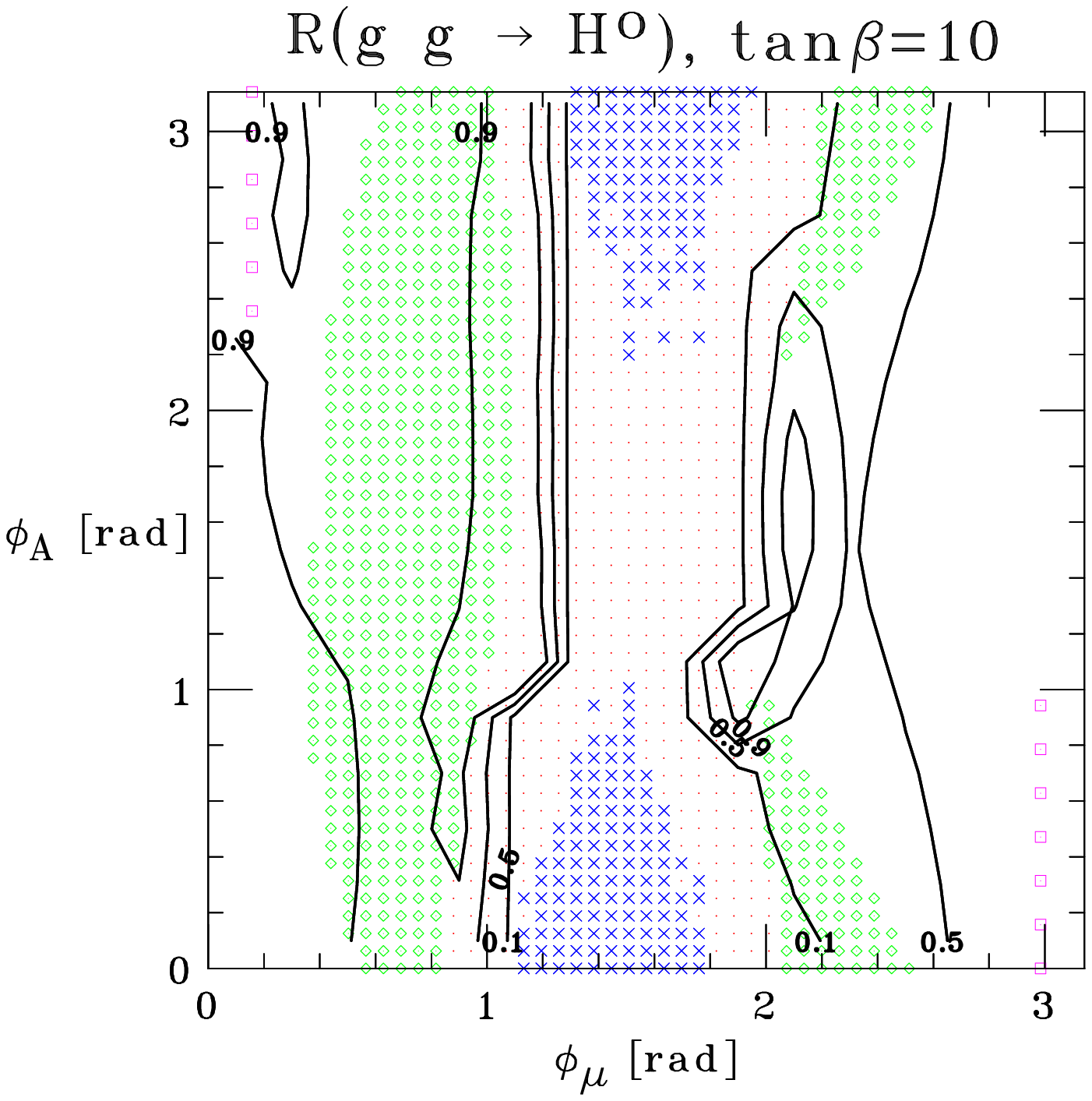,angle=90,height=3.25in}}}
\caption{Same as in Fig.~\ref{fig:h0} for the case $\Phi^0=H^0$.  }
\label{fig:H0}
\end{figure}

\begin{figure}
\centerline{\hbox{\epsfig{figure=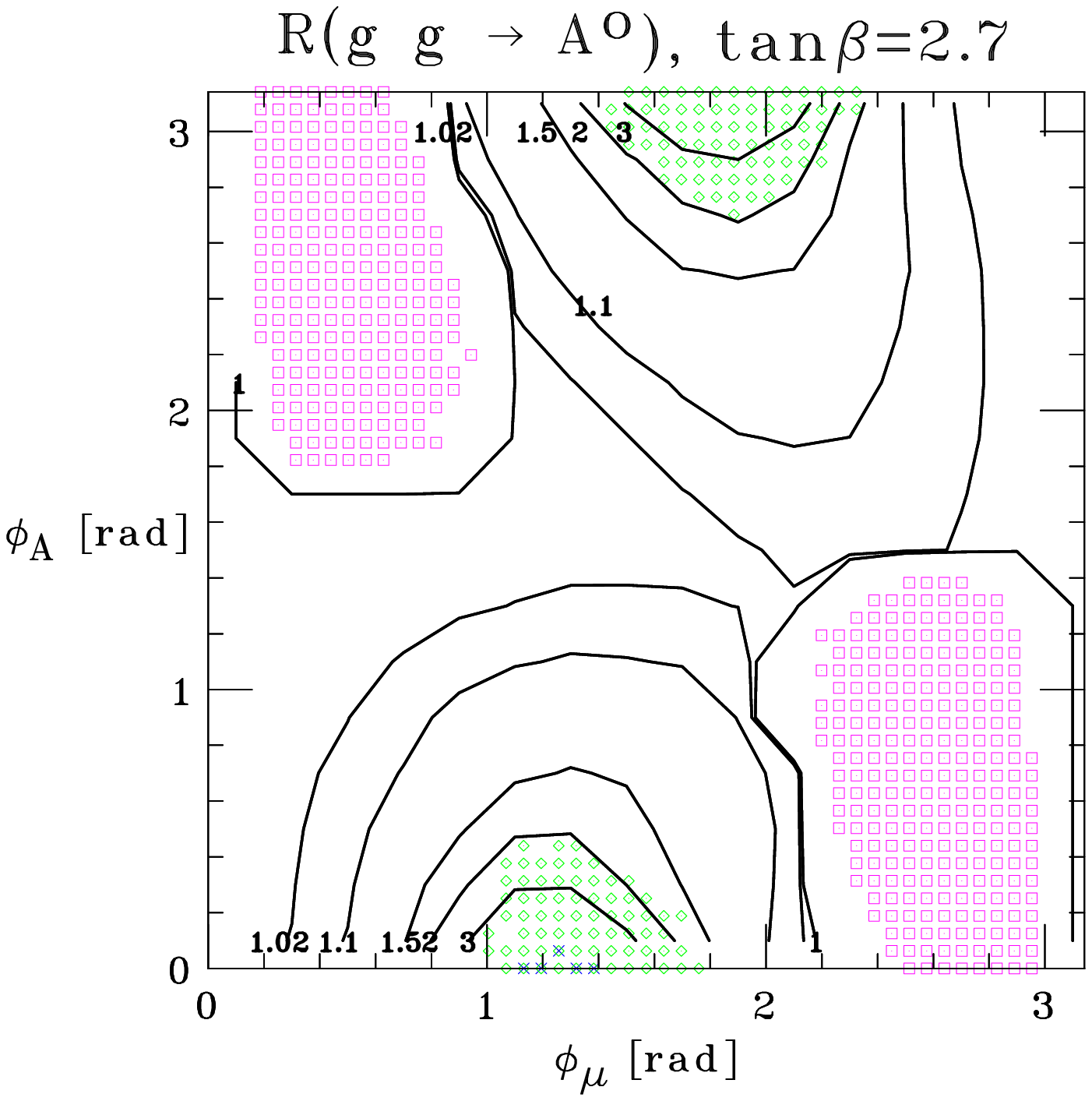,angle=90,height=3.25in}
\epsfig{figure=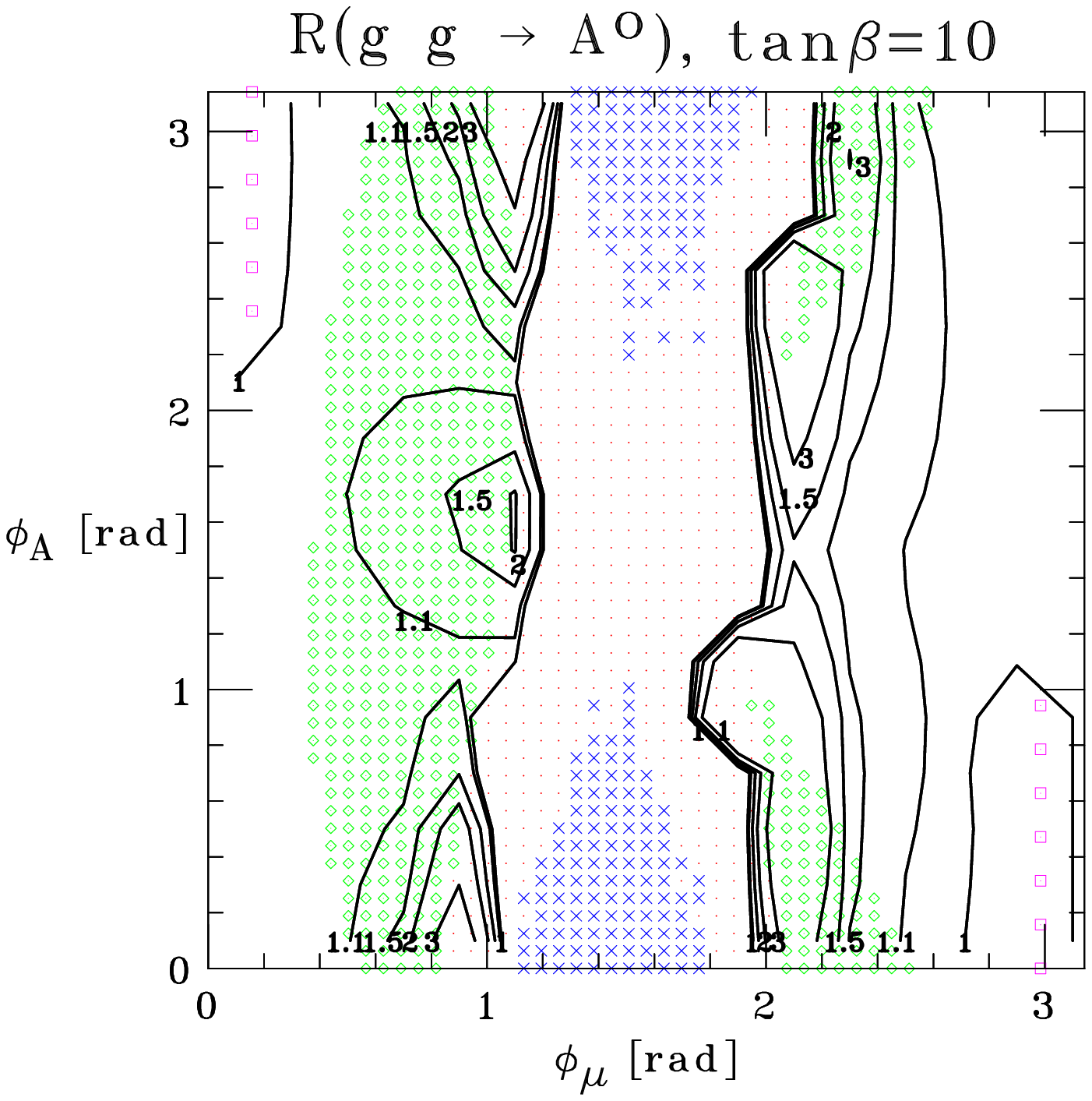,angle=90,height=3.25in}}}
\caption{Same as in Fig.~\ref{fig:h0} for the case $\Phi^0=A^0$.  }
\label{fig:A0}
\end{figure}

\begin{figure}
\centerline{\hbox{\epsfig{figure=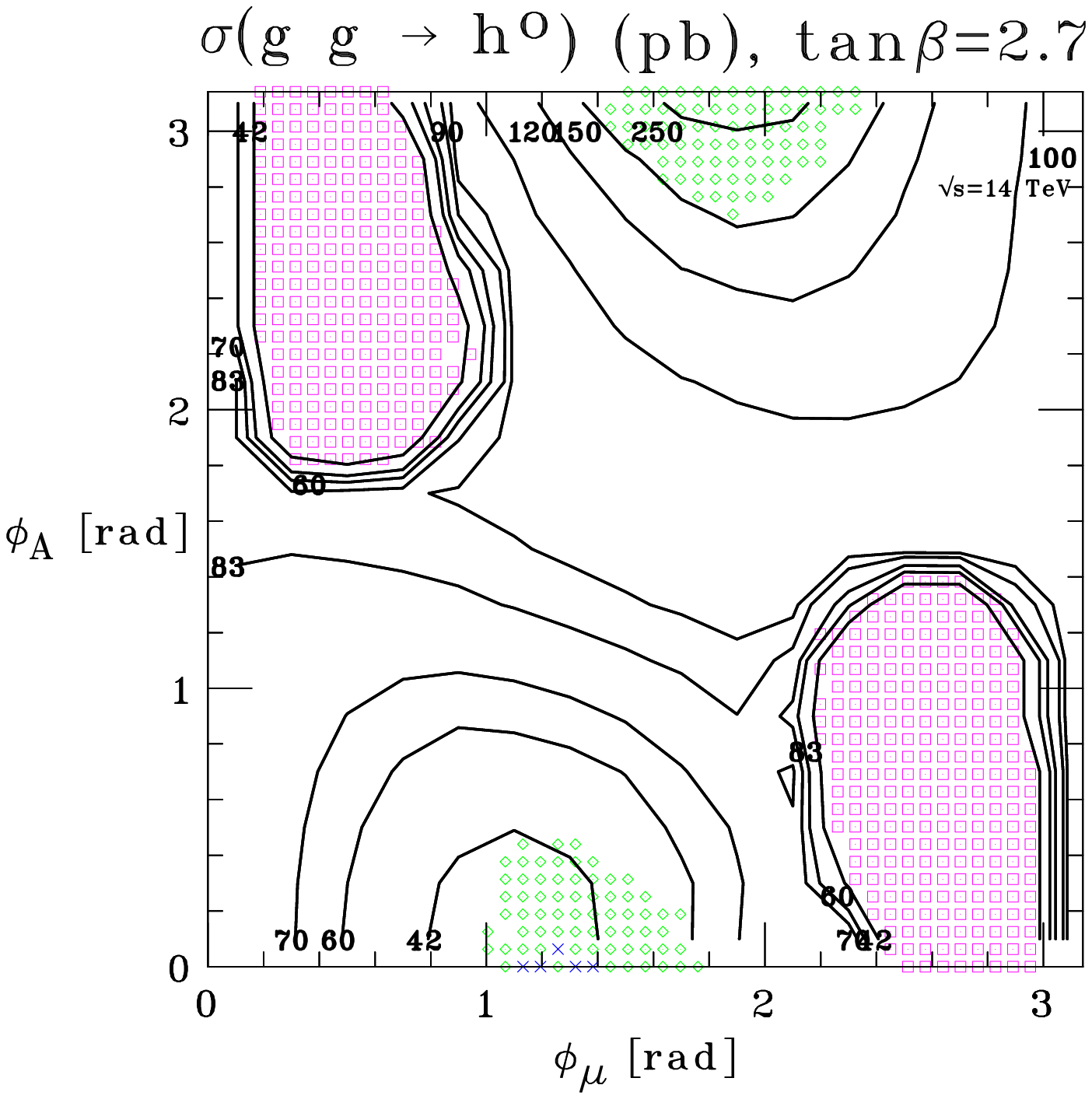,angle=90,height=3.25in}
\epsfig{figure=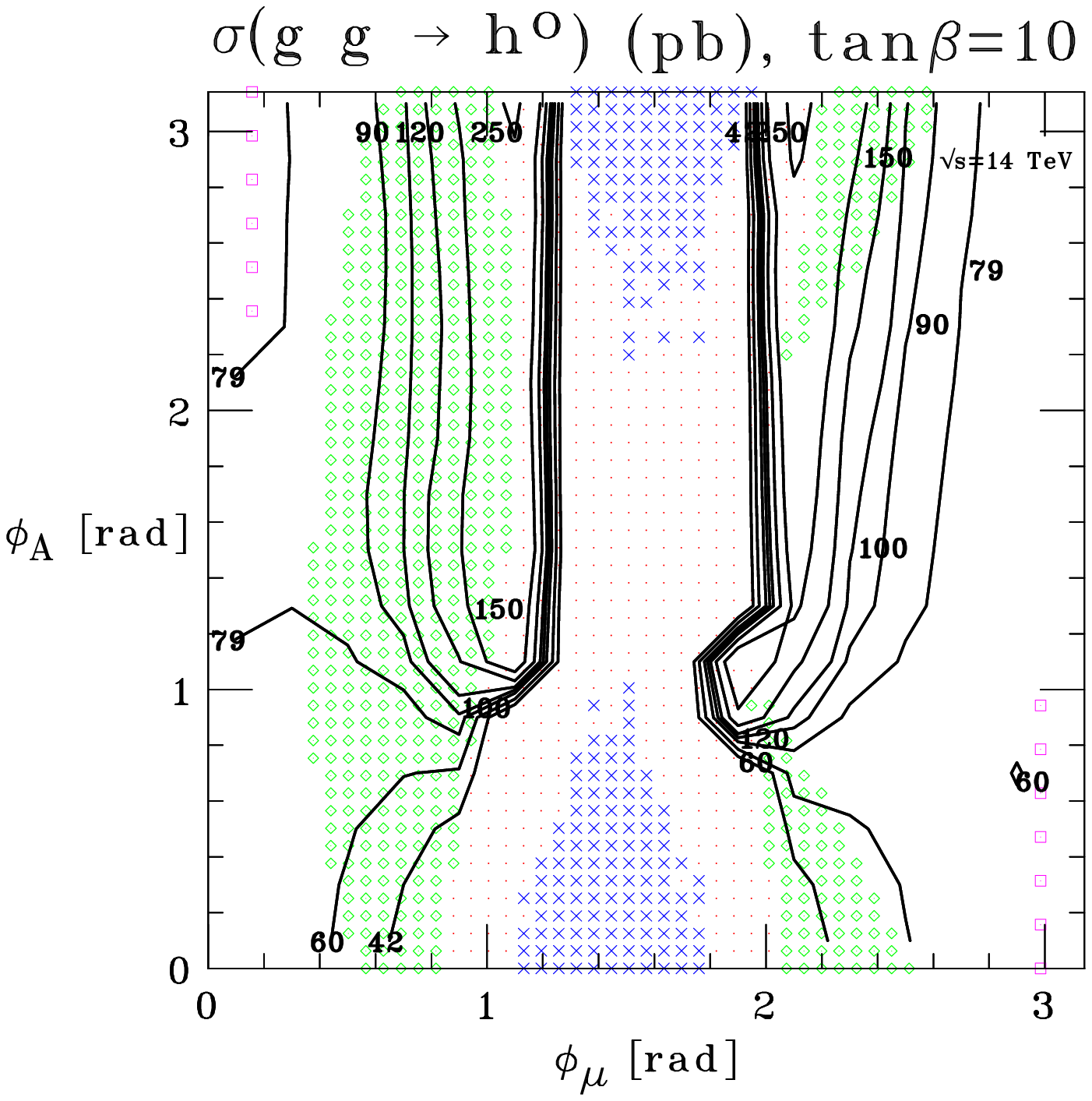,angle=90,height=3.25in}}}
\caption{Contour plots for the values of the NLO cross
section for $gg\to\Phi^0$ in the MSSM$^*$ at the LHC,
$\sigma^{\mathrm{MSSM}^*}_{\mathrm{NLO}}(gg\to\Phi^0)$,
 for the case $\Phi^0=h^0$, 
corresponding
to those of $|A|$ in Fig.~\ref{fig:A}, over the 
($\phi_\mu,\phi_A$) plane for small (left-hand plot) and 
large (right-hand plot) $\tan\beta$. The other MSSM parameters are as given in
Tab. I.}
\label{fig:sigmah0LHC}
\end{figure}

\begin{figure}
\centerline{\hbox{\epsfig{figure=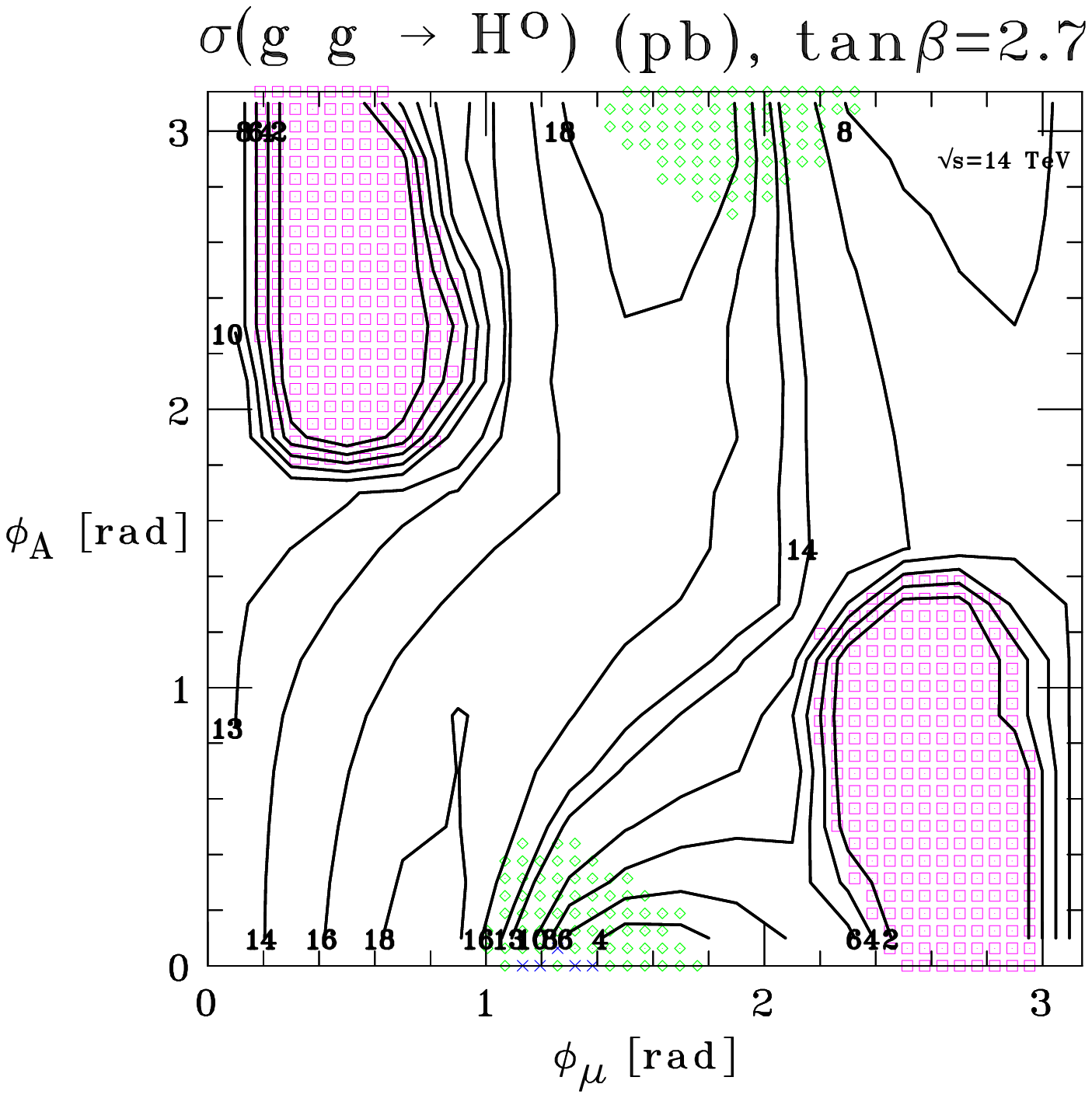,angle=90,height=3.25in}
\epsfig{figure=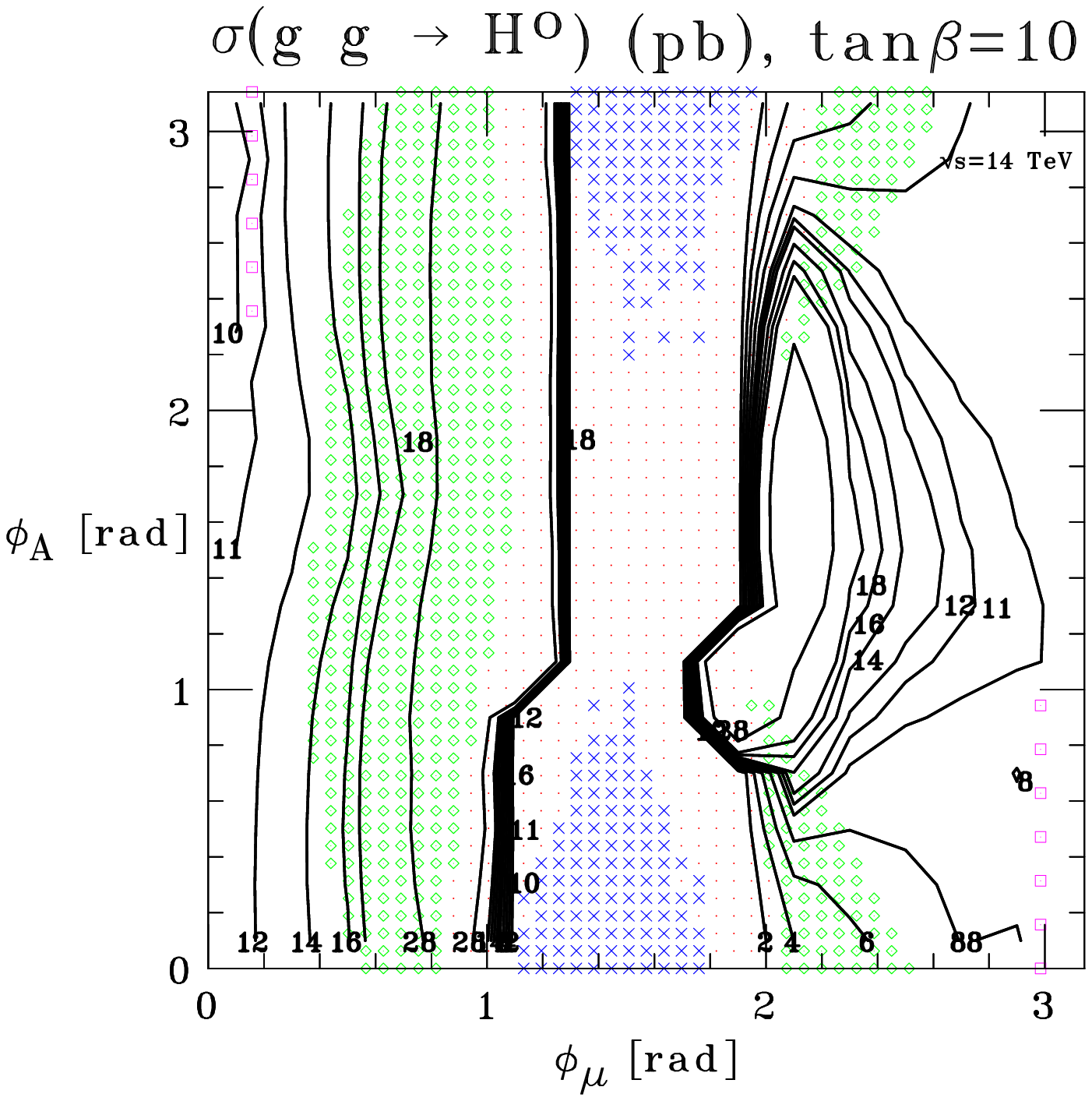,angle=90,height=3.25in}}}
\caption{Same as in Fig.~\ref{fig:sigmah0LHC} for the case $\Phi^0=H^0$.}
\label{fig:sigmaH0LHC}
\end{figure}

\begin{figure}
\centerline{\hbox{\epsfig{figure=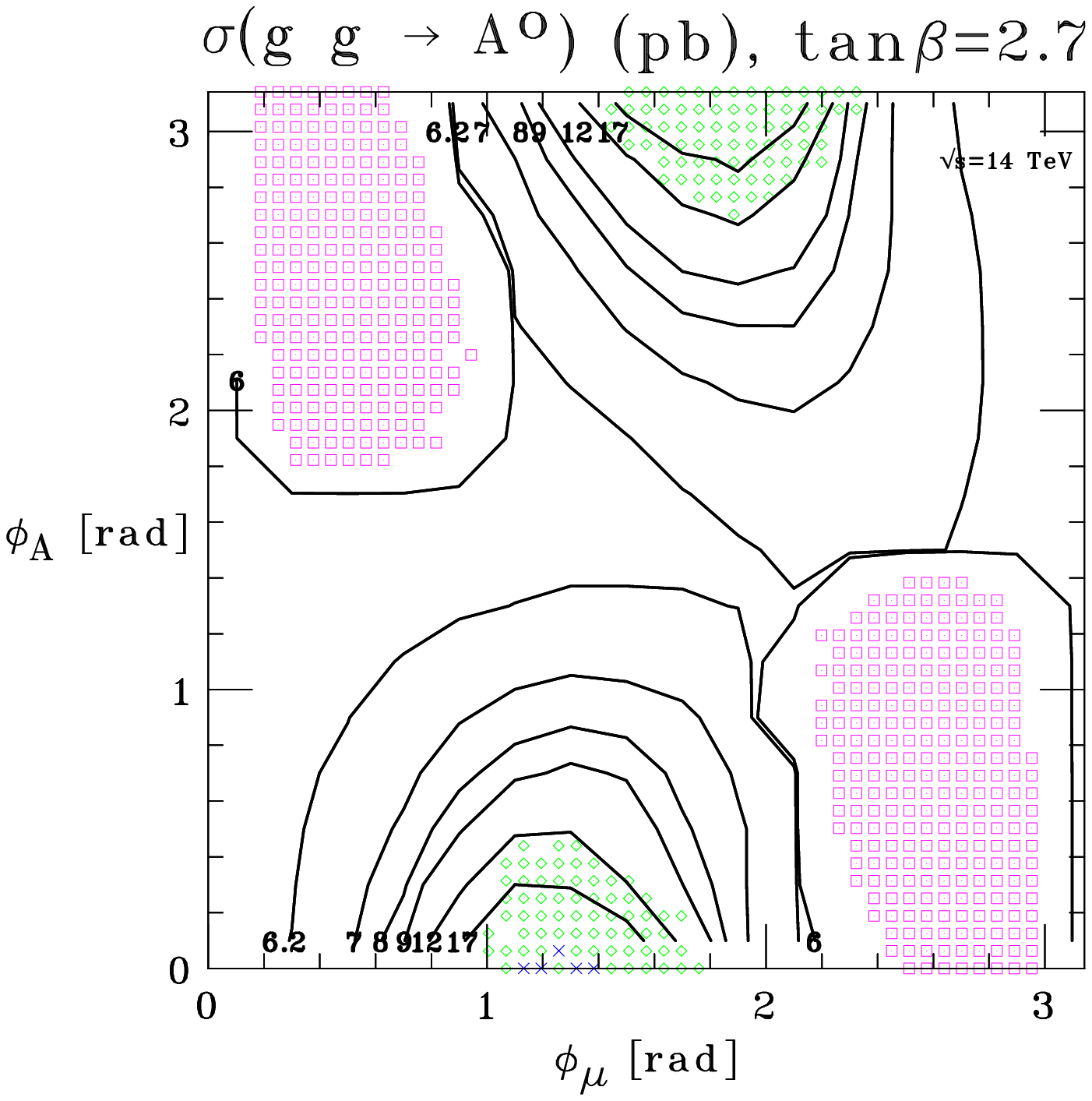,angle=90,height=3.25in}
\epsfig{figure=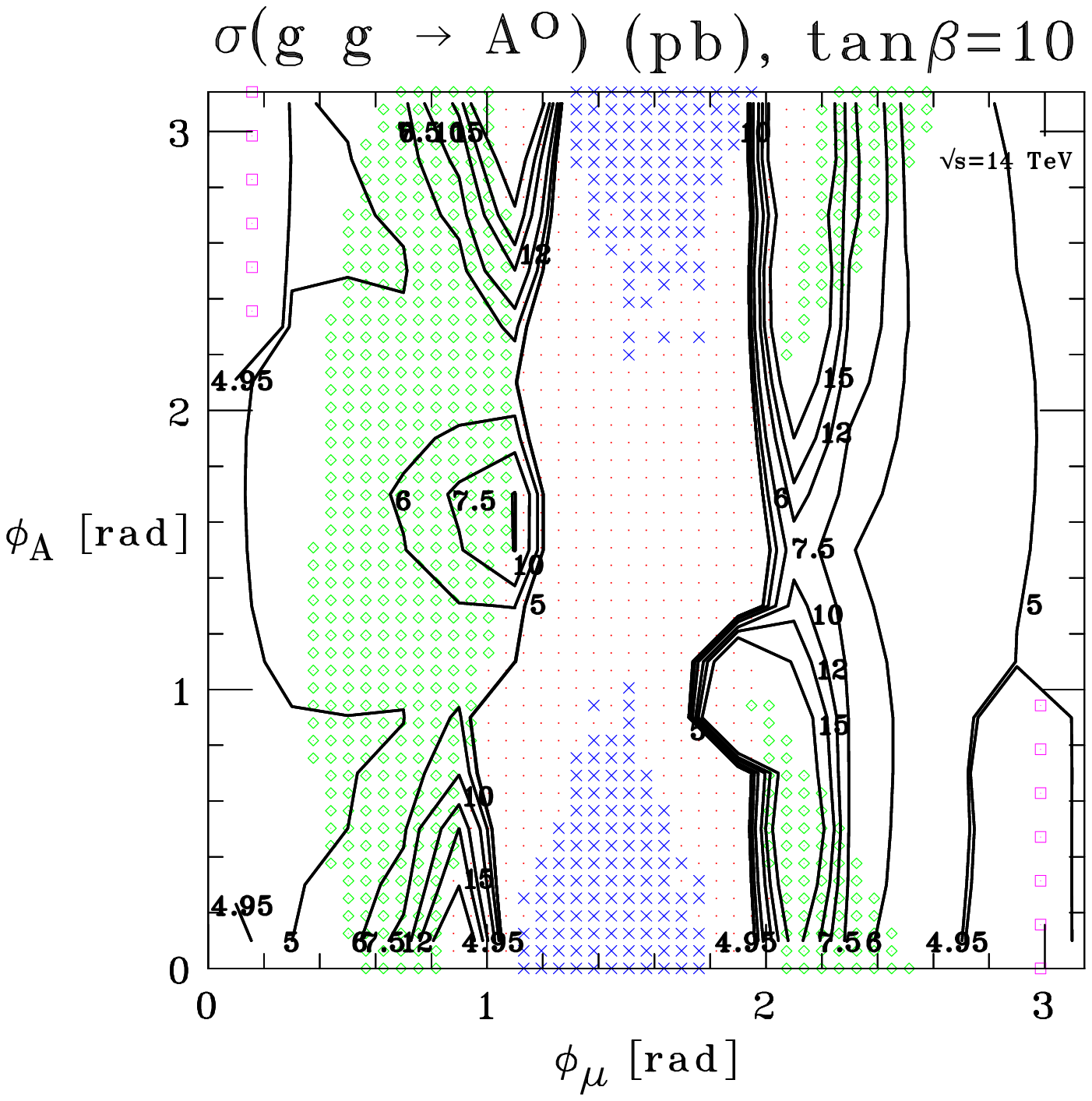,angle=90,height=3.25in}}}
\caption{Same as in Fig.~\ref{fig:sigmah0LHC} for the case $\Phi^0=A^0$.}
\label{fig:sigmaA0LHC}
\end{figure}

\begin{figure}
\centerline{\hbox{\epsfig{figure=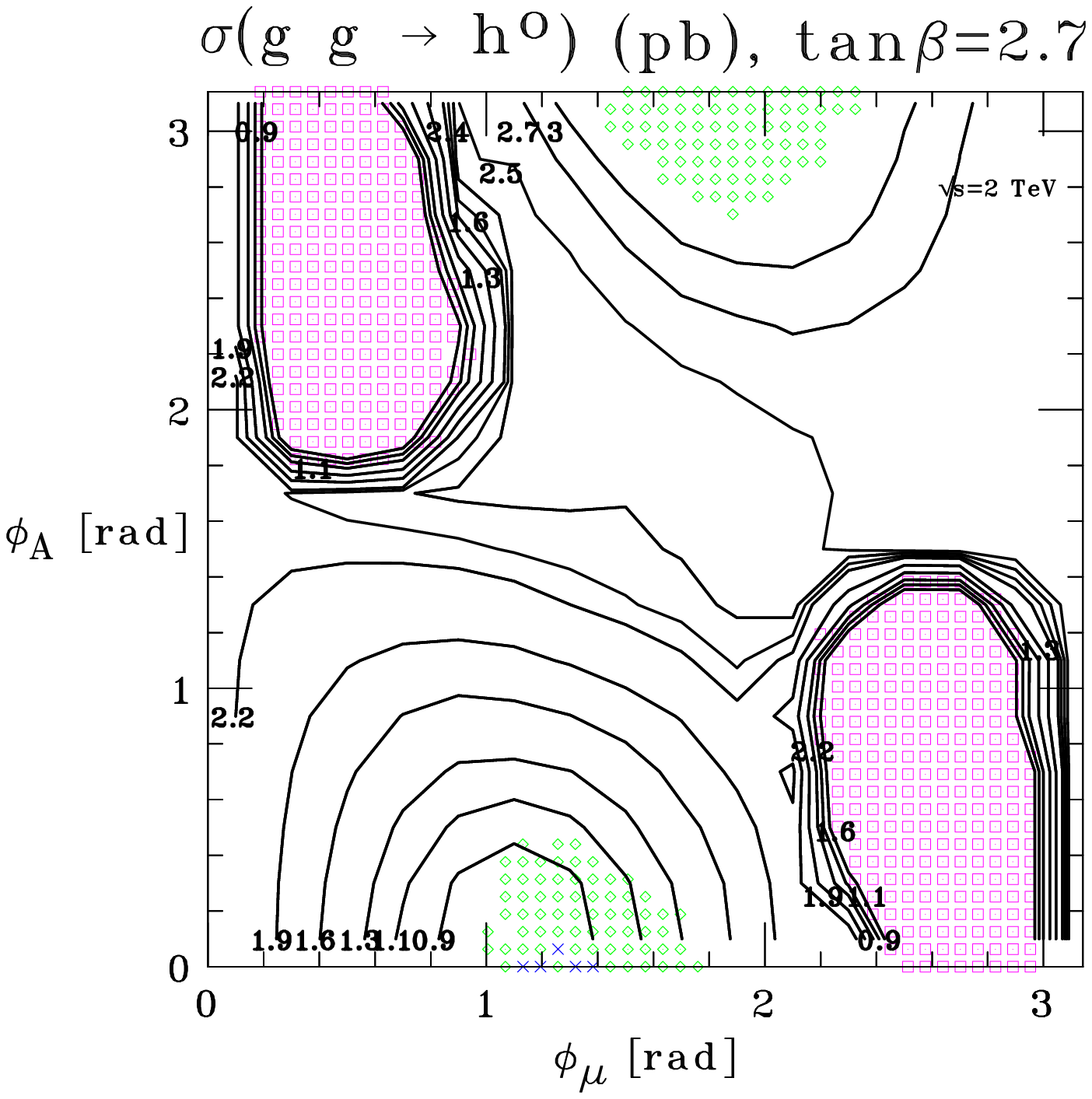,angle=90,height=3.25in}
\epsfig{figure=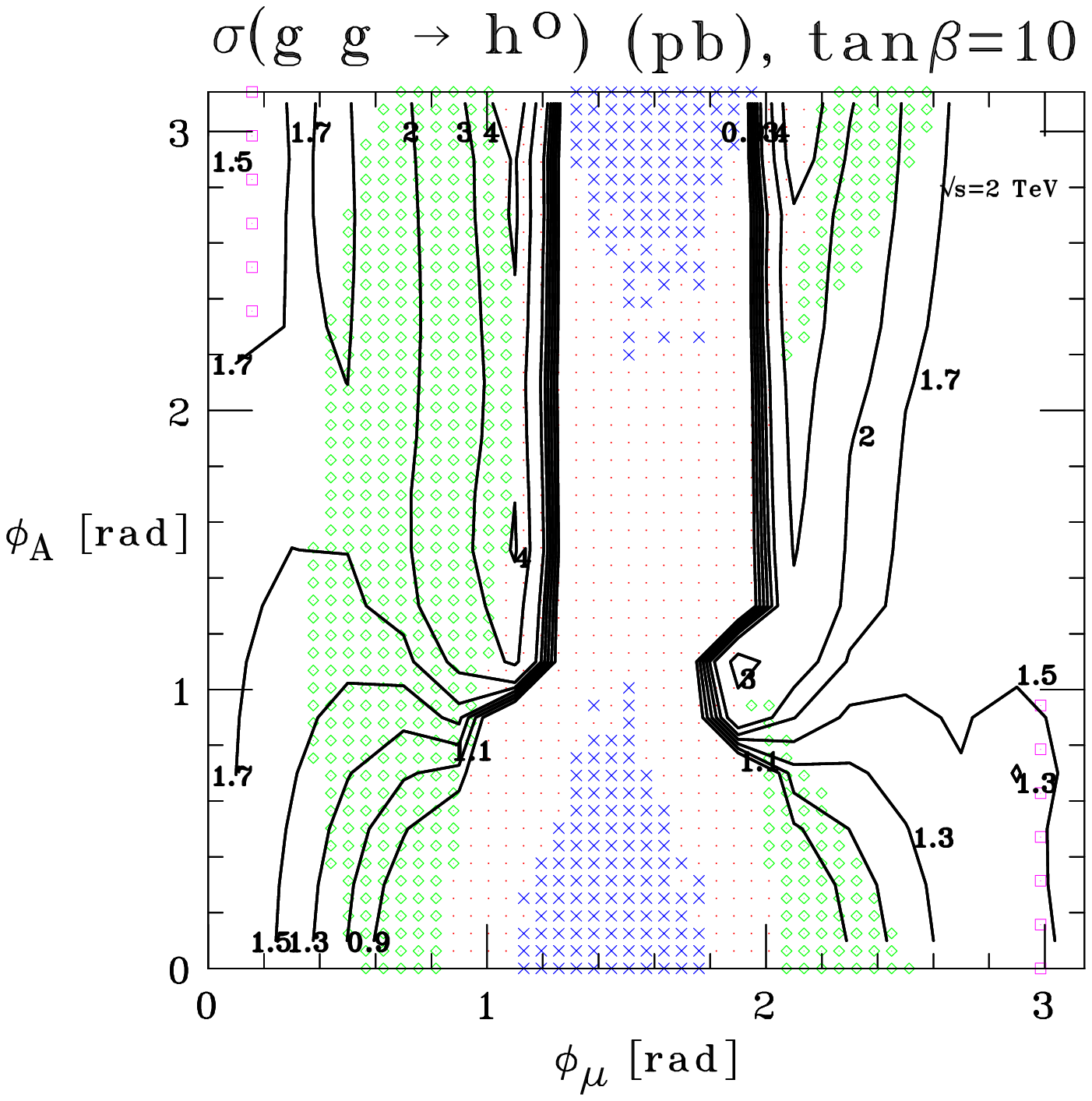,angle=90,height=3.25in}}}
\caption{Contour plots for the values of the NLO cross
section for $gg\to\Phi^0$ in the MSSM$^*$ at the Tevatron,
$\sigma^{\mathrm{MSSM}^*}_{\mathrm{NLO}}(gg\to\Phi^0)$,
 for the case $\Phi^0=h^0$, corresponding
to those  of $|A|$ in Fig.~\ref{fig:A}, over the 
($\phi_\mu,\phi_A$) plane for small (left-hand plot) and 
large (right-hand plot) $\tan\beta$. The other MSSM parameters are as given in
Tab. I.}
\label{fig:sigmah0TEV}
\end{figure}

\begin{figure}
\centerline{\hbox{\epsfig{figure=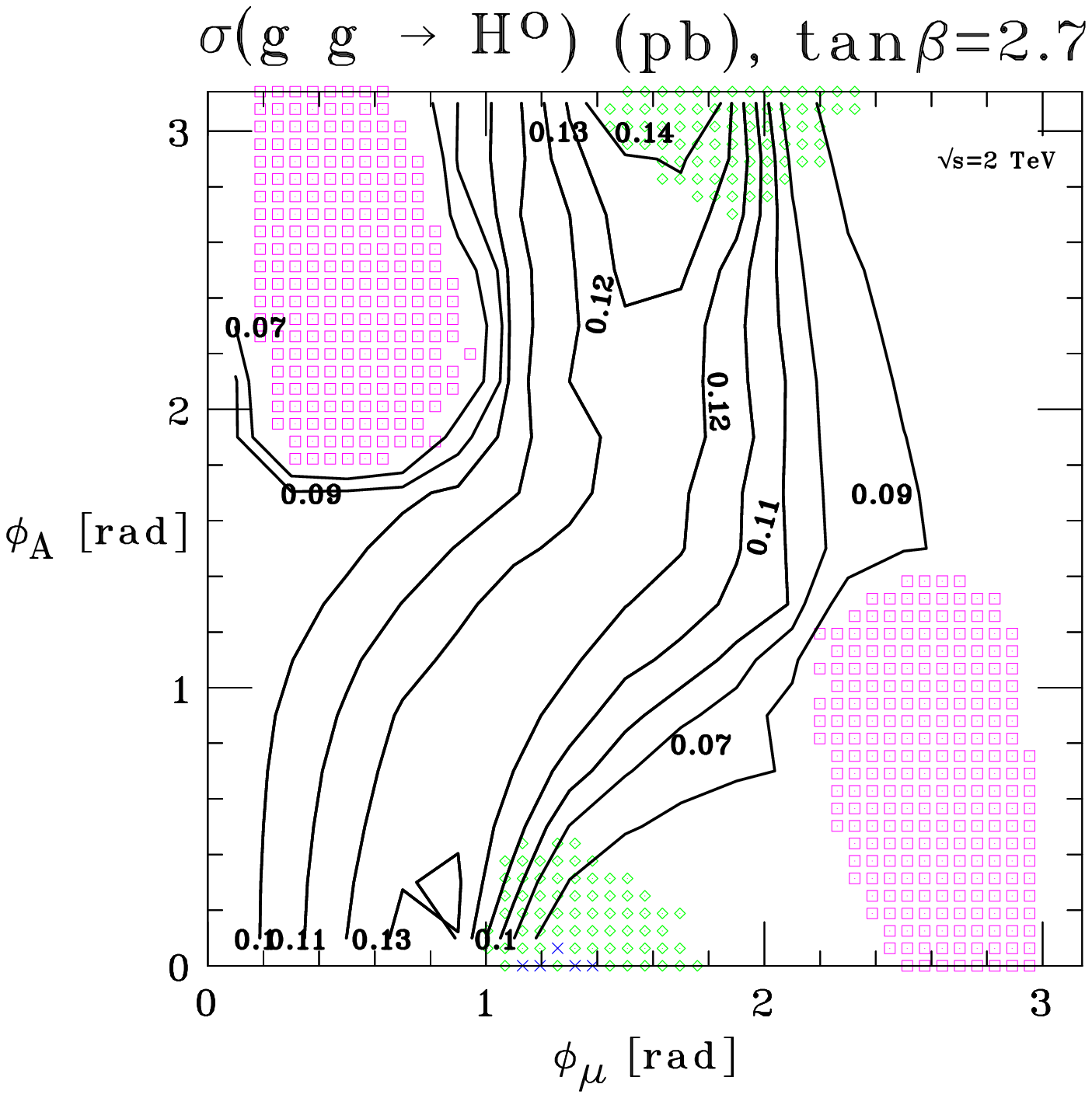,angle=90,height=3.25in}
\epsfig{figure=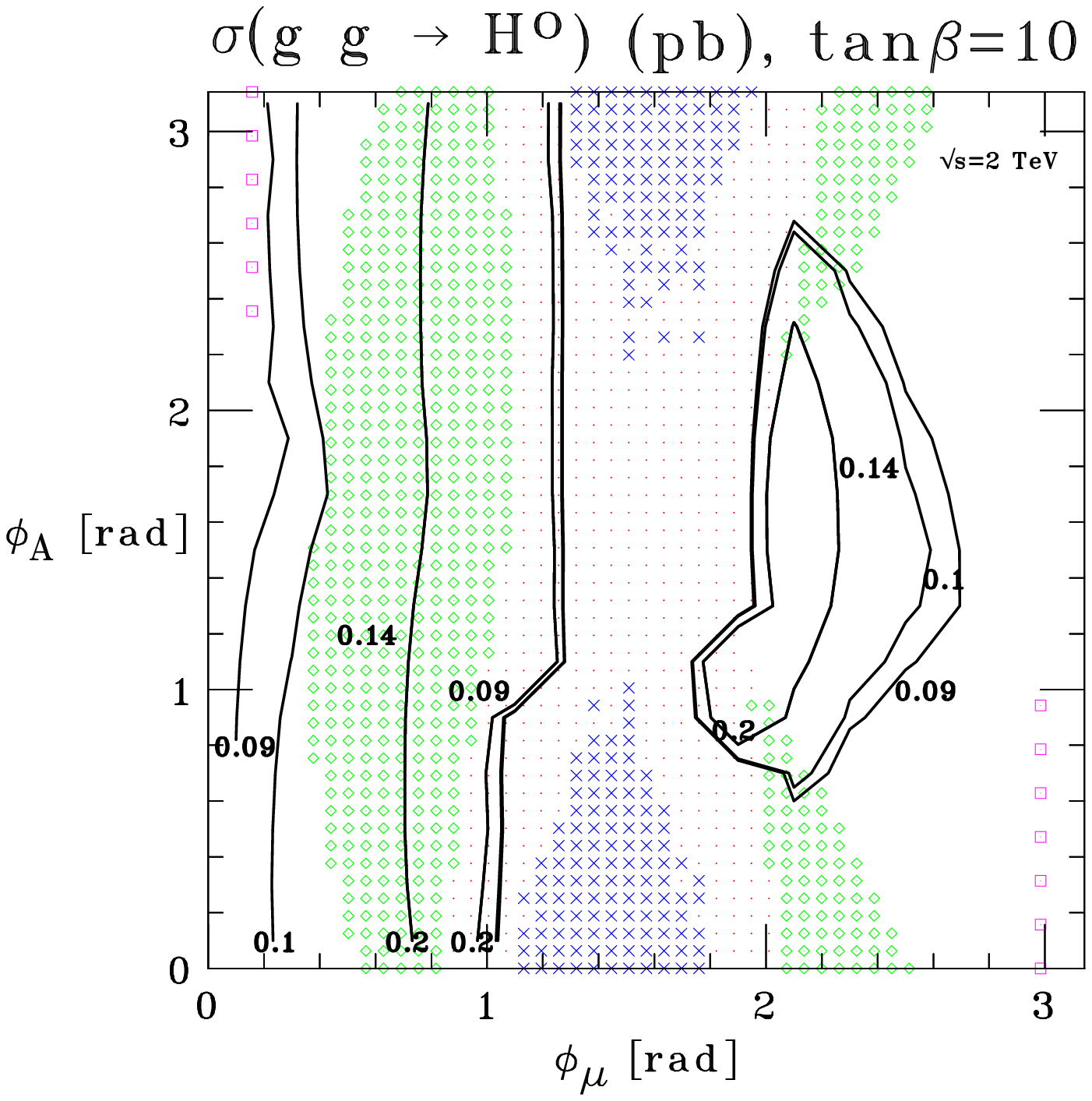,angle=90,height=3.25in}}}
\caption{Same as in Fig.~\ref{fig:sigmah0TEV} for the case $\Phi^0=H^0$.}
\label{fig:sigmaH0TEV}
\end{figure}

\begin{figure}
\centerline{\hbox{\epsfig{figure=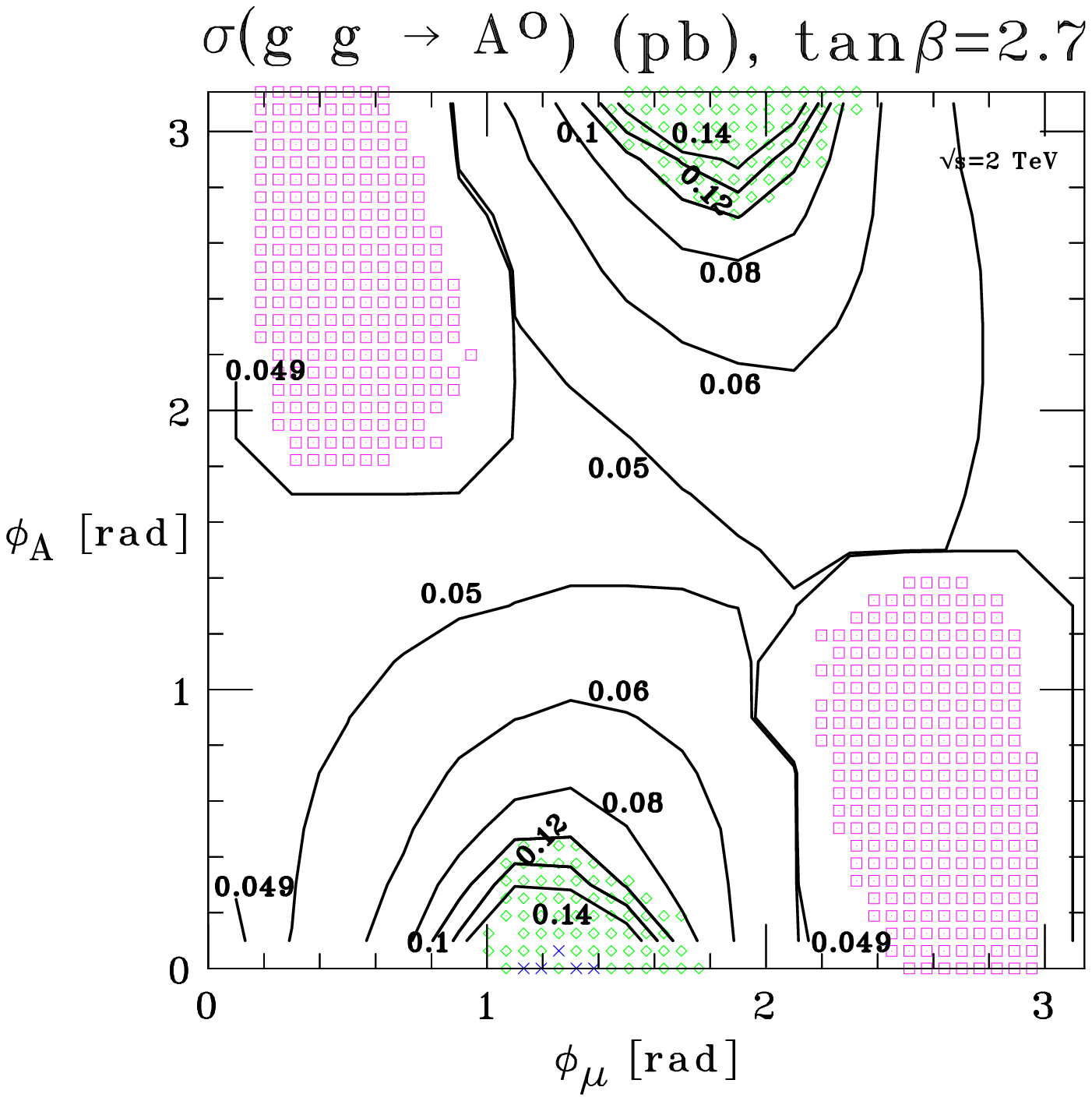,angle=90,height=3.25in}
\epsfig{figure=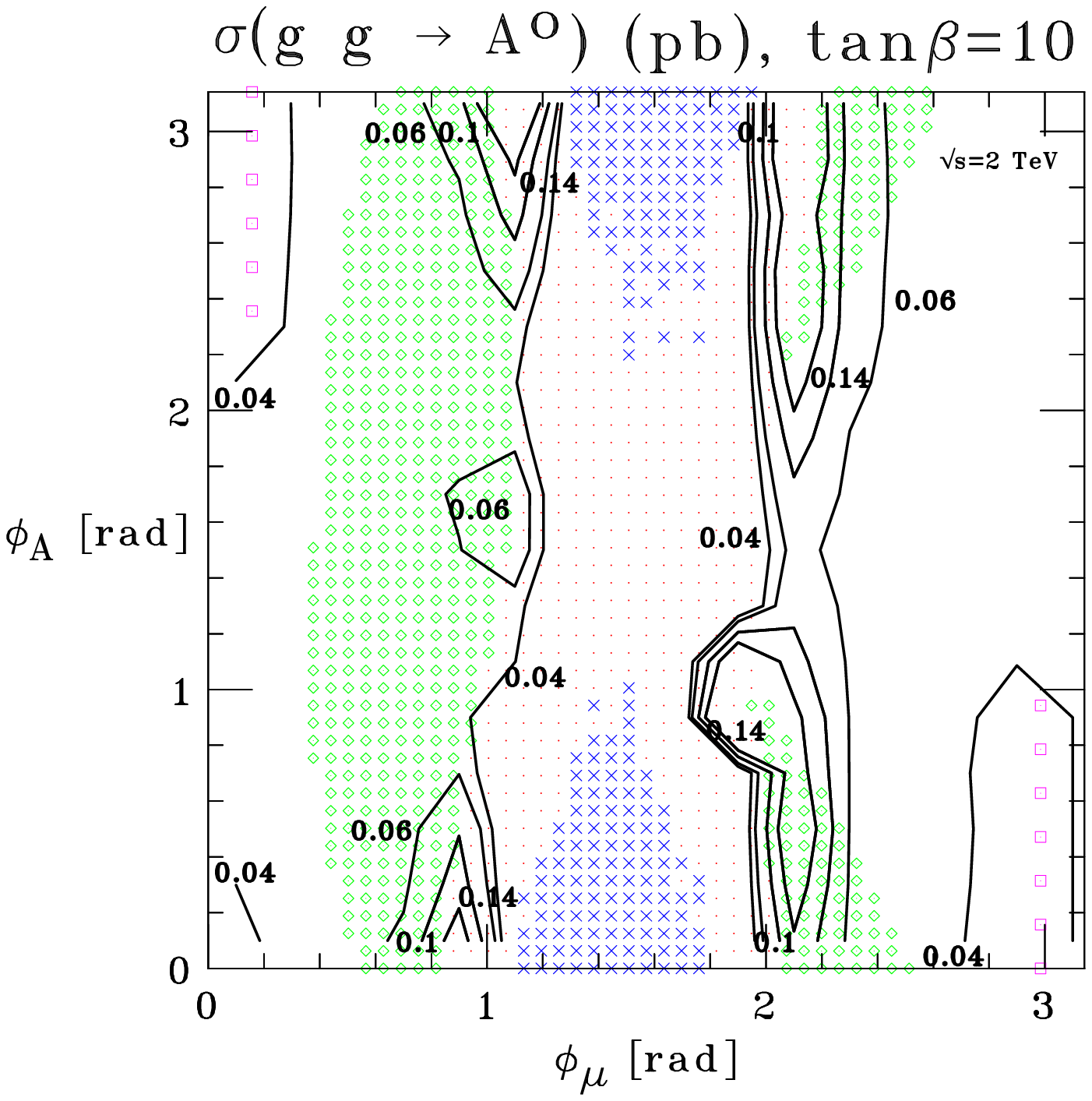,angle=90,height=3.25in}}}
\caption{Same as in Fig.~\ref{fig:sigmah0TEV} for the case $\Phi^0=A^0$.}
\label{fig:sigmaA0TEV}
\end{figure}

\end{document}